\documentclass[10pt]{IEEEtran}
\usepackage[utf8]{inputenc}
\usepackage{amsmath}
\usepackage{amsfonts}
\usepackage{amssymb}
\usepackage{mathtools}
\usepackage{amsthm}
\usepackage{fontenc}
\usepackage{graphicx}
\usepackage{balance}
\usepackage{booktabs}
\usepackage{subfigure}
\usepackage{tabulary}
\usepackage{booktabs}
\usepackage[justification=centering]{caption}
\usepackage{siunitx} 
\usepackage{comment}
\usepackage{cite}
\usepackage{algorithmic}
\usepackage{algorithm}
\usepackage[algo2e]{algorithm2e} 
\usepackage{hyperref}

\newtheorem{Prob}{\textbf{Problem}}

\newtheorem{Def}{Definition}
\SetKw{KwBy}{by}
\usepackage[colorinlistoftodos]{todonotes}
\usepackage{caption}
\newcommand{\highlighttext}[1] {#1}

\title{Joint Computing, Pushing, and Caching Optimization for Mobile Edge Computing Networks via Soft Actor-Critic Learning}
\author{Xiangyu Gao,~\IEEEmembership{Student Member,~IEEE}, Yaping Sun, Hao Chen, Xiaodong Xu, Shuguang Cui,~\IEEEmembership{Fellow,~IEEE}

\thanks{X.~Gao is with the Department of Electrical and Computer Engineering, University of Washington, Seattle, WA, USA. (email: xygao@uw.edu)}
\thanks{Y.~Sun and H.~Chen are with the Department of Broadband Communication, Peng Cheng Laboratory, Shenzhen 518000, China. (email: \{sunyp, chenh03\}@pcl.ac.cn)}
\thanks{X.~Xu is  with the Beijing University of Posts and Telecommunications, Beijing 100876, China, and affiliated with the Department of Broadband Communication, Peng Cheng Laboratory, Shenzhen 518000, China. (email: xuxiaodong@bupt.edu.cn)}
\thanks{S.~Cui is with the School of Science and Engineering (SSE) and \textcolor{black}{the Future Network of Intelligent Institute (FNii)}, the Chinese University of Hong Kong (Shenzhen), Shenzhen 518172, China. S.~Cui is also with Shenzhen Research Institute of Big Data, Shenzhen 518172, China, and affiliated with the Department of Broadband Communication, Peng Cheng Laboratory, Shenzhen 518000, China (email: shuguangcui@cuhk.edu.cn).}
\thanks{Corresponding Author: Yaping Sun.}
}

\begin{document}

\maketitle

\begin{abstract}
Mobile edge computing (MEC) networks bring computing and storage capabilities closer to edge devices, which reduces latency and improves network performance. However, to further reduce transmission and computation costs while satisfying user-perceived quality of experience, a joint optimization in computing, pushing, and caching is needed. In this paper, we formulate the joint-design problem in MEC networks as an infinite-horizon discounted-cost Markov decision process and solve it using a deep reinforcement learning (DRL)-based framework that enables the dynamic orchestration of computing, pushing, and caching. Through the deep networks embedded in the DRL structure, our framework can implicitly predict user future requests and push or cache the appropriate content to effectively enhance system performance. One issue we encountered when considering three functions collectively is the curse of dimensionality for the action space. To address it, we relaxed the discrete action space into a continuous space and then adopted soft actor-critic learning to solve the optimization problem, followed by utilizing a vector quantization method to obtain the desired discrete action. Additionally, an action correction method was proposed to compress the action space further and accelerate the convergence. Our simulations under the setting of a general single-user, single-server MEC network with dynamic transmission link quality demonstrate that the proposed framework effectively decreases transmission bandwidth and computing cost by proactively pushing data on future demand to users and jointly optimizing the three functions. We also conduct extensive parameter tuning analysis, which shows that our approach outperforms the baselines under various parameter settings.  
\end{abstract}


\section{Introduction}

Recent advancements in smart mobile devices have enabled various emerging applications such as virtual / augmented reality (VR/AR) \cite{vrsun}, online gaming, and autonomous driving \cite{gao2019experiments, gao2021perception, xiangyu2022deformable}. These applications require ultra-high communication and computation requirements, making it challenging for mobile operators to minimize communication and computation costs while ensuring the user-perceived quality of experience \cite{whitepaper}. In response to these challenges, the mobile edge computing network has emerged as a promising solution by pushing caching and computing resources to access points, base stations (BSs), and mobile devices at the wireless network edge \cite{vrsun}.

\subsection{Prior Art: Caching, Pushing, and Computing Design}
Caching can significantly improve bandwidth utilization by placing popular content closer to users for future reuse, leveraging the high degree of asynchronous content reuse in mobile traffic \cite{maddah}. Caching policies can be classified into two types, \textit{static caching} and \textit{dynamic caching}, depending on whether the cached contents remain unchanged or are dynamically updated. Static caching policies are generally determined based on the content popularity distribution, and the cache state remains unchanged for a relatively long time \cite{maddah,sundelay}. In \cite{sundelay}, a collaborative content caching scheme among base stations (BSs) in cache-enabled multi-cell cooperative networks is considered to minimize the average request delay, formulated by the stochastic request traffic modeling. Dynamic caching policies, on the other hand, involve updating the content placement from time to time by using instantaneous user request information, such as the least recently used (LRU) and least frequently used (LFU) policy \cite{10.1145/505696.505701}. However, since most caching policies are reactive operations and do not consider proactive pushing, the system performance can be further improved. 

Joint pushing and caching can indeed improve system performance by proactively transmitting contents during low traffic periods to satisfy future user demands. Several studies have explored this approach, such as \cite{sunpush} which considers a multi-user wireless network with multicast opportunities to minimize current and future transmission costs, and \cite{weidelay} which uses content request delay information to predict a user's request time for certain content items and maximize effective throughput. Additionally, \cite{codedpush} builds on RDI to characterize asynchronous user requests and proposes a coded joint pushing and caching method to minimize network traffic load with low complexity. In \cite{energy1}, a continuous-time optimization problem is formulated to determine optimal transmission and caching policies for small cell and Device-to-Device networks with known user demands in advance. However, existing joint pushing and caching policies only consider content delivery and have not taken into account the computation part, which limits their applicability to modern mobile traffic services such as mobile VR delivery.

To effectively serve modern mobile traffic, the joint design of caching, computing, and communication (3C) has been receiving increasing attention. One direction of 3C research focuses on the joint utilization of cache and computing resources at MEC servers to minimize transmission latency \cite{latency1,latency2} and energy consumption \cite{energy3c}. Another direction of 3C research considers the joint utilization of 3C resources at mobile devices to minimize communication costs in both single-user scenarios \cite{vrsun,single2} and multiple-user scenarios \cite{gainsun}. However, these joint 3C designs consider only static caching, where the cache states remain unchanged and do not take into account the benefits of pushing. Therefore, the system performance can be further improved through dynamic caching policies.

\subsection{DRL-based Systems}

Recent advances in deep learning (DL) have enabled the development of novel approaches for complicated classification and detection tasks \cite{ramp, 9765320}, as well as the solving of complex optimization problems that traditional methods may not be effective or efficient at handling \cite{9110932, 8513863, 8984310, 10.1007/s11036-018-1177-x, ZHAO2020101184, 8771176, gao2021mimosar, gao2023static}. Among all popular DL models, reinforcement learning (RL) \cite{9519528, haarnoja2018soft, mnih2015human} has been widely used in scheduling and optimization problems, such as transportation and resource allocation \cite{9247169, 10.1145/3389400.3389404, 9348485}, by learning an optimal policy for the agent to take actions that maximize a reward signal. By using RL, an agent can learn from experience and adapt its behavior over time to achieve the best possible outcomes. For example, in \cite{9110932}, a hierarchical RL algorithm is proposed to solve the joint optimization of pushing and caching in a multi-access edge computing network with multiuser and multicast data. The objective is to maximize bandwidth utilization and decrease the total quantity of data transmitted. In \cite{8513863}, the actor-critic RL framework is utilized to solve the joint optimization of caching, computation offloading, and radio resource allocation in the fog-enabled Internet of Things (IoTs), with the aim of minimizing the average end-to-end delay. In \cite{8984310}, Ning \textit{et al.} develop an intent-based traffic control system that utilizes DRL for the 5G-envisioned Internet of Connected Vehicles, which can dynamically orchestrate edge computing and content caching to improve the profits of mobile network operators. Furthermore, in \cite{10.1007/s11036-018-1177-x}, a distributed DL-based offloading algorithm is proposed, which uses multiple parallel deep neural networks to generate offloading decisions for MEC networks, where multiple wireless devices choose to offload their computation tasks to an edge server. In \cite{ZHAO2020101184, 8771176}, Zhao \textit{et al.} and Huang \textit{et al.} devise MEC networks for IoTs by using DRL frameworks to make the offloading strategy for offloading some computational tasks from IoT users to the computational access points or MEC server to reduce system latency and energy consumption. However, a common issue encountered when applying DRL-based systems to real-world optimization problems is the curse of dimensionality, which cannot be effectively and efficiently solved by general frameworks and optimization tools, especially for large-scale networks and tasks.

\subsection{Contributions}
To address the issues mentioned earlier, we propose a joint computing, pushing, and caching policy optimization framework in MEC networks\footnote{The code and sample data of this framework will be made open-source and available at
\href{https://github.com/Xiangyu-Gao/sac_joint_compute_push_cache}{\textit{https://github.com/Xiangyu-Gao/sac\_joint\_compute\_push\_cache}}}. Our contributions are as follows:
\begin{itemize}
\item We propose a model for the MEC network that computes its transmission and computation costs while taking into account computing, pushing, and caching actions. By representing system requests and their transition probabilities through a first-order F-state Markov chain, we formulate the joint optimization problem as an infinite-horizon discounted-cost Markov decision process with the dual objectives of reducing transmission and computation costs. Solving this problem requires dynamically optimizing the computing, pushing, and caching decisions over time to achieve the best overall performance.

\item To address the curse of dimensionality in the joint optimization problem, we implemented a continuous-space DRL approach known as soft actor-critic (SAC) learning \cite{haarnoja2018soft}. Unlike classic discrete-space DRL algorithms, such as deep Q-learning \cite{mnih2015human}, which rely on the Q-networks with a size linearly increased with the action space, SAC only requires learning the Gaussian-format Q-functions \cite{haarnoja2018soft}. As a result, SAC significantly reduces the number of parameters that need to be learned in a neural network. However, this does introduce the challenge of having an output action in continuous space that cannot be directly utilized. Therefore, we have designed an action quantization and correction algorithm that allows us to tailor SAC to our discrete optimization problem. Furthermore, the SAC algorithm is known for its stability and ease of convergence \cite{haarnoja2018soft}. 

\item We present simulation results with various system parameters under the setting of a general single-user, single-server MEC network to demonstrate the effectiveness of the proposed SAC algorithm. Our results show that by considering the joint optimization of computing, pushing, and caching, the performance of the MEC network can be significantly improved in terms of lower computation cost and reduced transmission cost. Moreover, our approach outperforms baseline methods that consider only a subset of these functions, demonstrating the benefits of the joint optimization. 
\end{itemize}

\subsection{Outline}
The paper is organized as follows:
Section II outlines the system model for the MEC network. Section III formulates the joint policy optimization problem. Section IV presents the utilization of SAC in optimization. Section V covers implementation details and evaluation results. Section VI concludes the paper.

\section{System Model}
\begin{figure}
\centering
\includegraphics[width=0.40\textwidth]{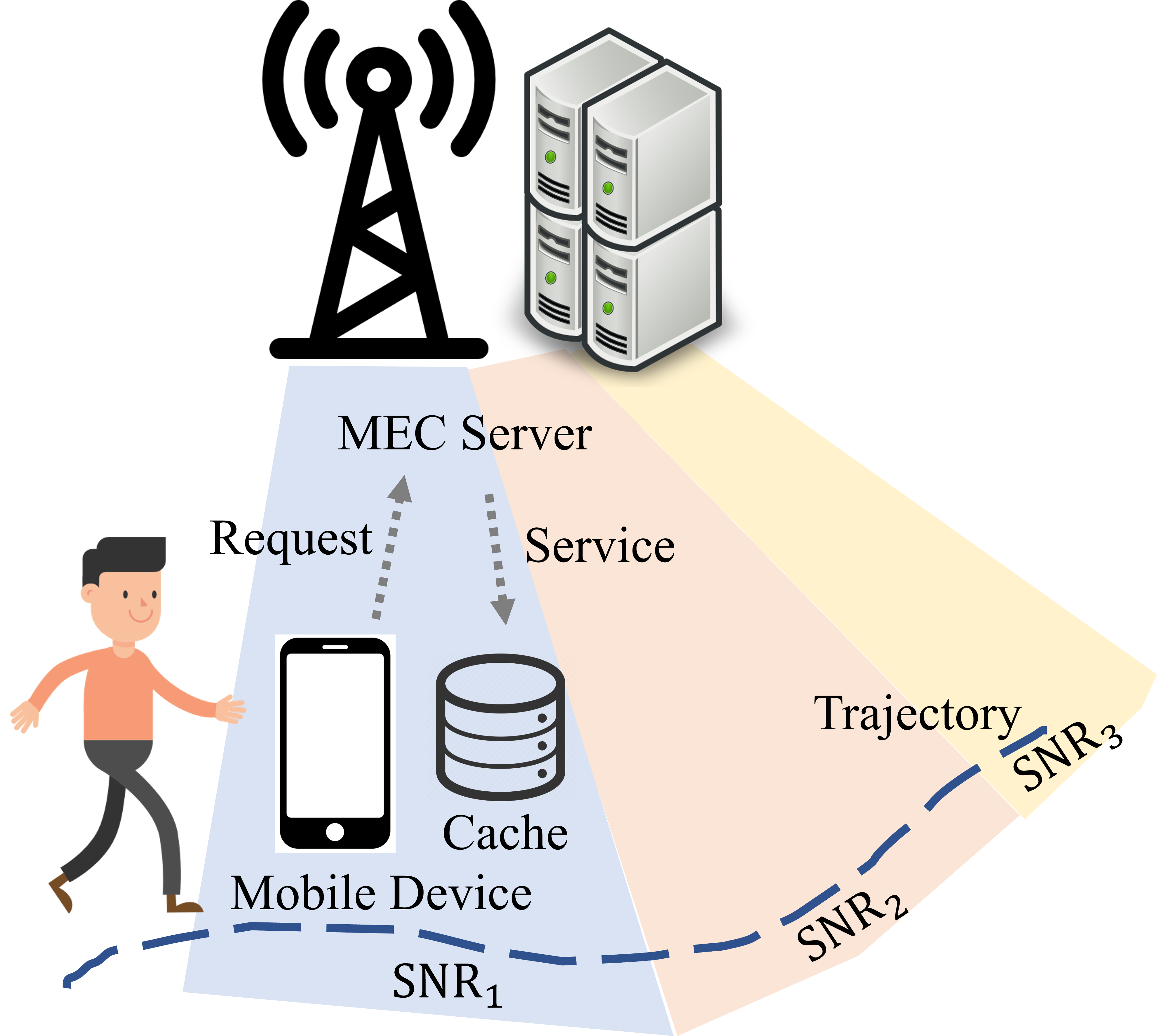}
  \caption{Illustration of MEC network with single MEC server and single mobile device. The mobile device is assumed to be moving at a small speed, such as an iPhone being carried by an individual. The channel quality for communication between the mobile device and the MEC server is modeled as an SNR, which may change over time due to the movement of the mobile device.}
  \label{fig:sys}
\end{figure}

Without loss of generality, we begin by considering a simple mobile edge network consisting of one MEC server and one mobile device, as shown in Fig.~\ref{fig:sys}. \highlighttext{The system model can be extended to the multi-user scenario by summing the objective functions of multiple users and considering the restrictions of the total communication and computing resources}. The MEC server has a large cache size, sufficient to proactively store the input and output data of all tasks requested by the mobile device. In contrast, the cache size of the mobile device is limited to a capacity denoted as $C$ (in bits). The mobile device is equipped with multi-core computing capabilities, each with a computation frequency $f_D$ (in cycles/s), and the number of computing cores is assumed to be $M$. The system operates over an infinite time horizon, with time slotted and indexed by $t=0,1,2,\cdots$, each with a fixed length of $\tau$ seconds. At the start of each time slot, the mobile device submits one task request, which is delay-intolerant and must be served before the end of the slot. \highlighttext{The tasks are categorized as delay-intolerant due to the critical importance of upholding optimal user experience and ensuring high-quality service for applications such as AR/VR, real-time communication, and streaming applications which are notably sensitive to delays.} Due to the mobility of the device, the data transmission rate for the link between the mobile device and the server may vary over time. To model this dynamic effect, we adopt the signal-to-noise ratio (SNR) following the Shannon theory, which measures the quality of the link. The system is designed to optimize the joint pushing, caching, and computing functions to minimize the computation and transmission costs of the network while ensuring timely and efficient task execution. 

\subsection{Task Model}
Assuming that the mobile device will request a total of $F$ tasks, we define the task set $\mathcal{F}$ as $\mathcal{F} \overset{\Delta}{=} \{1,2, \dots,f,\dots, F\}$. Each task $f\in \mathcal{F}$ is characterized by a $4$-item tuple $\Big\{I_f\ (\text{in\ bits}),\ O_f\  (\text{in\ bits}),\ w_f \ (\text{in cycles/bit}),\ \tau\ (\text{in seconds})\Big\}$. Specifically, $I_f$ represents the size of the input data generated from the Internet which can be cached. \highlighttext{$O_f$ represents the size of the output data after the computation is completed\footnote{In many systems, the actual output size might not be known beforehand, especially for computational tasks that involve dynamic data processing. It's possible that we could use historical data or estimations based on the characteristics of the input data and the computation process.}.} $w_f$ and $\tau$ denote the required computation cycles per bit and the maximum service latency, respectively.

\subsection{System State}
\subsubsection{Request State}
At each time slot $t$, the mobile device submits a single task request. \highlighttext{The request state at time $t$ is denoted by $A(t) \in \mathcal{F}$ representing the requested task, where $A(t)=f$ signifies that task $f$ in set $\mathcal{F}$ is being requested by the mobile device. The size of $\mathcal{F}$ is $F$. To model the evolution of requested tasks and their transition probabilities, we employed a first-order F-state Markov chain \cite{psounis2004modeling, 9110932}, referred to as ${A(t):t=0,1,2,\cdots}$. In this context, each state within the Markov chain corresponds to a distinct task, and the total number of tasks is assumed to be $F$. The choice to use a first-order Markov chain is rooted in its assumption that the probability of transitioning to a particular state is solely dependent on the current state.} The probability of transitioning to state $j \in \mathcal{F}$ at time slot $t+1$, given that the request state at time slot $t$ is $i\in \mathcal{F}$, is represented by $\Pr[A(t+1)=j | A(t)=i]$. It is assumed that ${A(t)}$ is time-homogeneous. We denote the transition probability matrix of ${A(t)}$ with $\textbf{Q}\triangleq\big(q_{i,j}\big)_{i\in\mathcal{F},j \in\mathcal{F}}$, where $q_{i,j}\triangleq \Pr \left[A(t+1)=j|A(t)=i\right]$. \highlighttext{Moreover, we focus our attention on an irreducible Markov chain to reflect the idea that any state in the system can be reached from any other state with a non-zero probability.} We denote the limiting distribution of ${A(t)}$ with $\textbf{p} \triangleq (p_{f})_{f\in \mathcal{F}}$. Here, $p_{f} \triangleq \lim_{t \to \infty} \Pr[A(t)=f]$, and it should be noted that $p_{f} = \sum_{i\in \mathcal{F}} p_{i}q_{i,f}$ for all $f\in \mathcal{F}$.

\subsubsection{Cache State} 
Let $S^I_f(t) \in \{0,1\}$ \highlighttext{denote the indicator of the cache state of the input data} for task $f$ stored in the mobile device. Here, $S^I_f(t) = 1$ means that the input data for task $f$ is cached in the mobile device, while $S^I_f(t) = 0$ implies that the input data is not cached. Similarly, let $S^O_f(t) \in \{0,1\}$ \highlighttext{denote the indicator of the cache state} of the output data for task $f$ stored in the mobile device, where $S^O_f(t) = 1$ represents that the output data for task $f$ is cached in the mobile device, and $S^O_f(t) = 0$ implies that the output data is not cached. The cache size of the mobile device is denoted by $C$ (in bits). The cache size constraint is given by
\begin{equation}
\label{cachesize}
\sum_{f=1}^F I_{f}S^I_{f}(t) + O_fS^O_f(t) \leq C
\end{equation}
which enforces that the sum of the sizes of input and output data cached for all tasks in the mobile device cannot exceed the cache size.

We define the cache state of the mobile device at time slot $t$, denoted by $\mathbf{S}(t) \triangleq (S^I_{f}(t), S^O_f(t))_{f\in\mathcal{F}} \in \textcolor{black}{\mathcal{S}}$, where \textcolor{black}{$\mathcal{S} \triangleq \{(S^I_{f},S^O_f)_{f\in\mathcal{F}}\in\{0,1\}^F \times \{0,1\}^F: \sum_{f\in \mathcal{F}} I_fS^I_{f}+O_fS^O_f \leq C\}$} represents the cache state space of the mobile device. Here, $N_{\min} \triangleq \frac{ C}{\max_{f\in \mathcal{F}} {{I_f,O_f}}}$ and $N_{\max} \triangleq \frac{ C}{\min_{f\in \mathcal{F}} {{I_f,O_f}}}$ represent the lower and upper bounds, respectively, on the cardinality of $\mathcal{S}$. The cardinality of $\mathcal{S}$ is bounded by $\binom{F}{N_{\min}}$ and $\binom{F}{N_{\max}}$ from below and above, respectively.

\subsubsection{System State}
At time slot $t$, the system state consists of both system request state and system cache state, represented by $\textbf{X}(t)\triangleq \left(A(t),\textbf{S}(t)\right)\in$ \textcolor{black}{$\mathcal{F}\times \mathcal{S}$}, where \textcolor{black}{$\mathcal{F}\times \mathcal{S}$} represents the system state space. 


\subsection{System Action} \label{sec:sys_action}
\subsubsection{Reactive Computation Action}
At each time slot $t$, the reactive transmission bandwidth cost and the reactive computation energy cost are denoted as $B^R(t)$ and $E^R(t)$, respectively. The task request $A(t)$ is served based on the current system state $\textbf{X}(t)=\left(A(t),\textbf{S}(t)\right)$ as follows:
\begin{itemize}
    \item
    If the cache state $S_{A(t)}^O(t)$ is equal to 1, it indicates that the output of task $A(t)$ is already cached locally, hence it can be retrieved without the need for any transmission or computation. As a result, the delay is negligible, and both the reactive computation energy and transmission cost become zero.
    
    \item 
    Assuming that $S_{A(t)}^I(t)=1$ and $S_{A(t)}^O(t) = 0$, it is possible to compute the requested task $A(t)$ directly using the locally cached input data. Let us define $c_{R,f}(t)\in \left\{1,\cdots, M\right\}$ as the number of computation cores allocated for reactively processing task $f$ at time slot~$t$ on the mobile device. Consequently, we can set $c_{R,f}(t) =0$ for all $f\in \mathcal{F}\backslash A(t)$. To ensure that the requested task $A(t)$ is completed within $\tau$, we must have $\frac{I_{A(t)}w_{A(t)}}{\tau} \leq c_{R,A(t)}(t) f_D$.\footnote{We assume that $\frac{I_{f}w_{f}}{\tau} \textbf{1}(A(t)=f) \leq M f_D$, for feasibility, where $\textbf{1}(A(t)=f)$ is the indicator function that is equal to 1 if $A(t)=f$, and 0 otherwise, and $M$ is the maximum number of computation cores. This assumption holds for all $f \in \mathcal{F}$.} Here $I_{A(t)}$ and $w_{A(t)}$ denote the input size and the computational workload of task $A(t)$, respectively. We can calculate the energy consumed for computing one cycle with frequency $c_{R,f}(t)f_D$ on the mobile device as $\mu c_{R,f}^2(t) f_D^2$, where $\mu$ is the effective switched capacitance related to the chip architecture indicating the power efficiency of the CPU. Therefore, the reactive computation energy cost $E^R(t)$ is given by $\mu c_{R,A(t)}^2(t)f_D^2I_{A(t)}w_{A(t)}$, and the reactive transmission cost $B^R(t)$ is zero.
    
    \item 
    If $S_{A(t)}^I(t)=0$ and $S_{A(t)}^O(t) = 0$, the mobile device must download the input data of task $A(t)$ from the MEC server before computing it locally. Let $\text{SNR}(t)$ be the SNR value of the data transmission link at time slot $t$. The required latency can be expressed as $\frac{I_{A(t)}}{B^R(t)\log_2{(1+\text{SNR}(t))}}+\frac{I_{A(t)}w_{A(t)}}{c_{R,A(t)}(t)f_D}$, where $B^R(t)\log_2{(1+\text{SNR}(t))}$ is the channel capacity given by Shannon theory. To satisfy the latency constraint, i.e., $\frac{I_{A(t)}}{B^R(t)\log_2{(1+\text{SNR}(t))}}+\frac{I_{A(t)}w_{A(t)}}{c_{R,A(t)}(t)f_D} \leq \tau$, the minimum reactive transmission cost $B^R(t)$ is given by $\frac{I_{A(t)}}{\left(\tau - \frac{I_{A(t)}w_{A(t)}}{c_{R,A(t)}(t)f_D}\right)\log_2{(1+\text{SNR}(t))}}$.\footnote{\highlighttext{The steps of deriving $B^R(t)$ from the preceding latency constraint are as follows: First, we have $I_{A\left(t\right)}/\left(B^R\left(t\right)\log_2\left(1+\text{SNR}(t)\right)\right)\le\tau-I_{A\left(t\right)}w_{A\left(t\right)}/\left(c_{R,A\left(t\right)}f_D\right)$. Then, we can get $I_{A\left(t\right)}/B^R\left(t\right)\le\left(\tau-I_{A\left(t\right)}w_{A\left(t\right)}/\left(c_{R,A\left(t\right)}f_D\right)\right)\log_2\left(1+\text{SNR}(t)\right)$. Finally, we can get
    $B^R\left(t\right)\geq I_{A\left(t\right)}/\left(\left(\tau-I_{A\left(t\right)}w_{A\left(t\right)}/\left(c_{R,A\left(t\right)}f_D\right)\right)\log_2\left(1+\text{SNR}(t)\right)\right)$.}} The reactive computation energy cost $E^R(t)$ is given by $\mu c_{R,A(t)}^2(t) f_D^2I_{A(t)}w_{A(t)}$.
    
\end{itemize}

In summary, at time slot $t$, the reactive computation action $c_{R,f}(t)$ should satisfy 
\begin{align}\label{eq:reacom}
   c_{R,f}(t) \leq \textbf{1}(A(t) = f)\left(1-S_f^O(t)\right)M, \ \forall f \in \mathcal{F}, 
\end{align}
\noindent and the reactive transmission cost $B^R(t)$ is given by
\begin{align}\label{eq:reactivebandwidth}
B^R(t) = & \left(1-S_{A(t)}^I(t)\right)\left(1-S_{A(t)}^O(t)\right) \\
& \times \frac{I_{A(t)}}{\left(\tau - \frac{I_{A(t)}w_{A(t)}}{c_{R,A(t)}(t)f_D}\right)\log_2{(1+\text{SNR}(t))}}, 
\end{align}
and the reactive computation cost $E^R(t)$ is given by
\begin{equation}\label{eq:reactiveenergy}
    E^R(t) = \left(1-S_{A(t)}^O(t)\right)\mu c_{R,A(t)}^2(t)f_D^2I_{A(t)}w_{A(t)}. 
\end{equation} 

Let $\textbf{c}_R \triangleq (c_{R,f})_{f\in \mathcal{F}}\in \Pi_C^R(\textbf{X})$ denote the reactive computation action of the system, where $\Pi_C^R(\textbf{X}) \triangleq \left\{(c_{R,f})_{f\in \mathcal{F}} \in \left\{0,1,\cdots,M\right\}^F: (\ref{eq:reacom}) \right\}$ represents the decision space for reactive computation of the system under state $\textbf{X}$. It can be observed from Eq.~\eqref{eq:reacom} that the size of the reactive computation action space is $M+1$.

\subsubsection{Proactive Transmission or Pushing Action} 
Let $b_{f}(t) \in \{0,1\}$ denote the binary decision variable for task $f \in \mathcal{F}$, where $b_{f}(t)=1$ indicates that the remote input data of task $f$ is pushed to the mobile device, and $b_{f}(t) = 0$ otherwise. We assume that the pushed data is transmitted to the mobile device by the end of the time slot. To ensure compliance with the latency constraint, we enforce $\frac{\sum_{f=1}^FI_fb_{f}(t)}{\tau} \leq B^P(t)\log_2{(1+\text{SNR}(t))}$, where $B^P(t)$ denotes the proactive transmission bandwidth cost. Thus, the minimum proactive transmission cost can be expressed as:
\begin{equation}\label{eq:probandwidth}
    B^P(t) = \frac{\sum_{f=1}^F I_fb_{f}(t)}{\tau \log_2{(1+\text{SNR}(t))}}.
\end{equation}

In summary, the system pushing action under system state $\mathbf{b} \triangleq \left(b_{f}\right)_{f\in \mathcal{F}} \in \{0,1\}^F$. The size of the system pushing action space under system state $\textbf{X}$ is $2^F$

\subsubsection{Cache Update Action}
The cache state of each task $f\in \mathcal{F}$ is updated according to 
\begin{align}
    &S_f^I(t+1) = S_f^I(t) + \Delta s_f^I(t),\label{inputupdate}\\
    &S_f^O(t+1) = S_f^O(t) + \Delta s_f^O(t),\label{outputupdate}
\end{align}

\noindent where $\Delta s_f^I(t) \in \{-1,0,1\}$ and $\Delta s_f^O(t) \in \{-1,0,1\}$ denote the update action for the cache state of the input and output data of task $f$, respectively. Then, we have $\forall f \in \mathcal{F}$
\begin{align}
    & -S_f^I(t) \leq \Delta s_f^I(t) \leq \min \left\{b_f(t)+c_{R,f}(t), 1-S_f^I(t)\right\} \label{cachinp1} \\
    & -S_f^O(t) \leq  \Delta s_f^O(t) \leq \min \left\{c_{R,f}(t), 1-S_f^O(t)\right\},
    \label{cachoutp1}\\
    & \resizebox{.88\hsize}{!}{$\sum_{f=1}^F I_f\left(S_f^I(t)+\Delta s_f^I(t)\right)+O_f\left(S_f^O(t)+\Delta s_f^O(t)\right) \leq C$},\label{eq:caupsize}
\end{align}
where the left-hand side of Eq.~\eqref{cachinp1} specifies that the removal of the input of task $f$ from the mobile device is only possible if it has been previously cached. On the other hand, the right-hand side of Eq.~\eqref{cachinp1} indicates that the caching of the input of task $f$ into the mobile device is only allowed if it has not been cached before and if it is either proactively transmitted from the MEC server or reactively transmitted, i.e., if $b_f(t) = 1$ or $c_{R,f}(t) > 0$. Similarly, the left-hand side of Eq.~\eqref{cachoutp1} states that the output of task $f$ can only be removed from the mobile device if it has been previously cached. On the other hand, the right-hand side of Eq.~\eqref{cachoutp1} specifies that the caching of the output of task $f$ into the mobile device is only allowed if it has not been cached before and if it is reactively computed at the mobile device, i.e., if $c_{R,f}(t) > 0$. Finally, Eq.~\eqref{eq:caupsize} requires that the updated cache state complies with the cache size constraint.

In summary, let $\Delta \textbf{s} \triangleq \left(\Delta s_f^I,\Delta s_f^O\right){f\in \mathcal{F}} \in \Pi_{\Delta s}(\textbf{X})$ denote the system cache update action, where $\Pi_{\Delta s}(\textbf{X}) \triangleq \left\{\! \left(\Delta s_f^I,\Delta s_f^O\right)_{f\in \mathcal{F}}\! \in\! \{-1,0,1\}^F\! \times \!\{-1,0,1\}^F: \eqref{cachinp1}, \eqref{cachoutp1},\eqref{eq:caupsize}\right\}$. Here, $\Pi_{\Delta s}(\textbf{X})$ represents the system cache update action space for the given system state $\textbf{X}$, and it includes tuples of $\Delta s_f^I$ and $\Delta s_f^O$ for each task $f \in \mathcal{F}$, with values in $\{-1,0,1\}$ indicating whether to evict, retain, or cache a task's input and output data.

\subsubsection{System Action}
The system action at each time slot is a combination of three distinct actions: reactive computation, pushing, and cache update. This combination is represented as $\left(\textbf{c}_{R},\textbf{b},\Delta \textbf{s}\right)  \in \Pi(\textbf{X})$, where $\Pi(\textbf{X})$ is the system action space under the current system state $\textbf{X}$,  $\Pi(\textbf{X}) \triangleq \Pi_C^R(\textbf{X})\times \{0,1\}^F\times \Pi_{\Delta s}(\textbf{X})$. 

\subsection{System Cost}
At each time slot $t$, the overall system cost is a combination of two components, namely the transmission bandwidth cost and the computation energy cost. The transmission bandwidth cost consists of both proactive and reactive transmission costs and is given by 
\begin{equation}\label{bandwidth}
    B(t) = B^R(t) + B^P(t),
\end{equation}

\noindent where $B^R(t)$ is given in Eq.~\eqref{eq:reactivebandwidth} and $B^P(t)$ is given in Eq.~\eqref{eq:probandwidth}. he reactive computation cost contributes to the computation energy cost only and is given by
\begin{equation}\label{energy}
    E(t) = E^R(t),
\end{equation}
\noindent where $E^R(t)$ is given in Eq.~\eqref{eq:reactiveenergy}. 

 To strike a balance between communication and computation cost, the system cost at time slot $t$ is computed as the weighted sum of transmission bandwidth cost and computation energy cost, i.e., $B(t)+\lambda E(t)$, where $\lambda$ is a non-negative weighting factor.

\section{Problem Formulation}
Given an observed system state $\textbf{X}$, the joint reactive computing, transmission, and caching action, denoted as $\left(\textbf{c}_{R}, \textbf{b},\Delta \textbf{s}\right)$, is determined according to a policy defined as below. 
\begin{Def}[Stationary Joint Computing, Pushing and Caching Policy] A stationary joint computing, pushing, and caching policy $\pi$ is a mapping from system state $\textbf{X}$ to system action $\left(\textbf{c}_{R},\textbf{b},\Delta \textbf{s}\right)$, i.e., $\left(\textbf{c}_{R},\textbf{b},\Delta \textbf{s}\right)= \pi(\textbf{X}) \in \Pi(\textbf{X})$.
\end{Def}

From properties of $\{A(t)\}$ and $\{\textbf{S}(t)\}$, the induced system state process $\{\textbf{X}(t)\}$ under policy $\pi$ is a controlled Markov chain. The expected total discounted cost $ \phi(\pi)$ is given as:
\begin{align}
    \phi(\pi)\triangleq \limsup_{T\rightarrow \infty} \sum_{t=0}^{T-1}\gamma^{t}\mathbb{E}\left[B(t)+\lambda E(t)\right],
\end{align}
\highlighttext{where $T$ is the length of the request process, $\gamma$ is the discount factor, $B(t), E(t)$ are the transmission bandwidth cost and computation energy cost at time $t$, and $\lambda$ is the weight balancing two costs.}

In this paper, we aim to obtain optimal joint computing, pushing, and caching policy to minimize the sum of infinite horizon discounted system cost, i.e., minimize both the transmission and computation cost, as follows:

\begin{Prob}[Joint Computing, Pushing and Caching Policy Optimization]
    \begin{align}
        \phi^* \triangleq & \min_{\pi} \ \ \phi(\pi) \nonumber\\
        &\  s.t.\ \  \pi(\textbf{X}) \in \Pi(\textbf{X}),\ \  \forall   \textbf{X} \in \mathcal{F}\times \mathcal{S}. \nonumber
    \end{align}
\end{Prob}

\section{Soft Actor-Critic Learning}

\subsection{SAC System State and Action} \label{sec:SAC_action}

The system state $\mathbf{x}$ of SAC is designed the match the system state $\textbf{X}$ in the formulated problem, such that $\mathbf{x}=\textbf{X}=\left(A(t),\textbf{S}(t)\right)$, with a vector size of $2F+1$. 

The SAC algorithm is designed to solve continuous-action problems, whereas the required system action $\left(\textbf{c}_{R},\textbf{b},\Delta \textbf{s}\right)$ in the formulated problem is discrete. To address this issue, we define the system action of the SAC as the \textbf{\textit{continuous version}} of the formulated system action space. This continuous version is denoted as $\mathbf{a}=\left(\bar{\textbf{c}}_{R},\bar{\textbf{b}},\Delta \bar{\textbf{s}}\right) \in \bar{\Pi}(\textbf{X}) \triangleq \bar{\Pi}_C^R(\textbf{X})\times [0,1]^F\times \bar{\Pi}_{\Delta s}(\textbf{X})$. Here, $\bar{\Pi}_C^R(\textbf{X}) \triangleq \left\{(c_{R,f})_{f\in \mathcal{F}} \in \left[0,M\right]^F: (\ref{eq:reacom}) \right\}$, and $\bar{\Pi}_{\Delta s}(\textbf{X}) \triangleq \left\{ \left(\Delta s_f^I,\Delta s_f^O\right)_{f\in \mathcal{F}} \in [-1,1]^F \times [-1,1]^F: \eqref{cachinp1}, \eqref{cachoutp1}, \eqref{eq:caupsize}\right\}$. 

As $\bar{\textbf{c}}_{R}\triangleq \left\{(\bar{c}_{R,f})_{f\in \mathcal{F}}\right\}$ must always equal zero for $f\in \mathcal{F}{\setminus} A(t)$, the action space of SAC can be simplified by disregarding the computing cores for non-requested tasks. We can obtain the simplified form of action $\mathbf{a}$ as $\mathbf{a}=\left(\bar{c}_{A(t)},\bar{\textbf{b}},\Delta \bar{\textbf{s}}\right)$, with a vector size of $3F+1$.

\subsection{SAC Learning}
\label{sec:sac_learning}
SAC is an off-policy deep reinforcement learning method that maintains the advantages of entropy maximization and stability while offering sample-efficient learning \cite{haarnoja2018soft}. It operates on an actor-critic framework where the actor is responsible for maximizing expected reward while simultaneously maximizing entropy. The critic evaluates the effectiveness of the policy being followed.

A general form of maximum-entropy RL is given by:
\begin{equation}
    J(\pi)=\sum_{t=0}^T \mathbb{E}_{\left(\mathbf{x}_t, \mathbf{a}_t\right) \sim \rho_\pi}\left[r\left(\mathbf{x}_t, \mathbf{a}_t\right)+\alpha \mathcal{H}\left(\pi\left(\cdot \mid \mathbf{x}_t\right)\right)\right]
    \label{eq:ge_rl}
\end{equation}
where the temperature parameter $\alpha$ determines the relative importance of the entropy term against the reward $r$, and the entropy term is given by $\mathcal{H}\left(\pi\left(\cdot \mid \mathbf{x}_t\right)\right)=\mathbb{E}_{\mathbf{a}_t}\left[-\log \pi\left(\mathbf{a}_t \mid \mathbf{x}_t\right)\right]$.

The SAC algorithm is a policy iteration approach designed to solve the optimization problem in Eq.~\eqref{eq:ge_rl} \cite{haarnoja2018soft}. It comprises two essential components: soft Q-function $Q_{\theta}\left(\mathbf{x}_{t}, \mathbf{a}_{t}\right)$, and policy $\pi_{\phi}\left(\mathbf{a}_{t} \mid \mathbf{x}_{t}\right)$. To deal with the large continuous domains, neural networks are utilized to approximate these components, with the network parameters denoted by $\theta$ and $\phi$. For example, the policy is modeled as a Gaussian distribution with a fully connected network providing the mean and covariance value, and the Q-function is also approximated using a fully connected neural network. Following \cite{haarnoja2018soft}, the update rules for $\theta$ and $\phi$ are provided below.

The soft Q-function parameters can be trained to minimize the soft Bellman residual
\begin{equation}
    \label{eq:J_Q}
    \begin{aligned} 
    J_{Q}(\theta)=\mathbb{E}_{\left(\mathbf{x}_{t}, \mathbf{a}_{t}\right) \sim \mathcal{D}}\Bigl[ 
    & \frac{1}{2}\bigl(Q_{\theta}\left(\mathbf{x}_{t}, \mathbf{a}_{t}\right)-\bigl(r\left(\mathbf{x}_{t}, \mathbf{a}_{t}\right)+ \\
    & \gamma \mathbb{E}_{\mathbf{x}_{t+1} \sim p}\left[V_{\bar{\theta}}\left(\mathbf{x}_{t+1}\right)\right]\bigr)\bigr)^{2}\Bigr],
    \end{aligned}
\end{equation}
where $\mathcal{D}$ is the distribution of previously sampled states and actions, $p$ is the transition probability between states, and the value function $V_{\bar{\theta}}(\mathbf{x}_t)$ is implicitly parameterized through the soft Q-function parameters as follows
\begin{equation}
V_{\bar{\theta}}\left(\mathbf{x}_t\right)=\mathbb{E}_{\mathbf{a}_t \sim \pi}\left[Q_{\bar{\theta}}\left(\mathbf{x}_t, \mathbf{a}_t\right)-\alpha \log \pi\left(\mathbf{a}_t \mid \mathbf{x}_t\right)\right]
\end{equation}

The update makes use of a target soft Q-function $Q_{\bar{\theta}}$ with parameters $\bar{\theta}$ obtained as an exponentially moving average of the soft Q-function weights $\theta$, which helps stabilize training. The soft Bellman residual $J_{Q}(\theta)$ in Eq.~\eqref{eq:J_Q} can be optimized with stochastic gradients
\begin{equation}
    \label{eq:J_Q_grad}
    \begin{aligned} 
   \hat{\nabla}_\theta J_Q(\theta)= & \nabla_\theta Q_\theta\left(\mathbf{a}_t, \mathbf{x}_t\right)\Bigl(Q_\theta\left(\mathbf{x}_t, \mathbf{a}_t\right)-\bigl(r\left(\mathbf{x}_t, \mathbf{a}_t\right) + \\ &
  \gamma\left(Q_{\bar{\theta}}\left(\mathbf{x}_{t+1}, \mathbf{a}_{t+1}\right)-\alpha \log \left(\pi_\phi\left(\mathbf{a}_{t+1} \mid \mathbf{x}_{t+1}\right)\right)\right)\bigr)\Bigr).
    \end{aligned}
\end{equation}

The policy parameters $\phi$ can be learned by directly minimizing the expected KL divergence in
\begin{equation}
    \label{eq:J_Pi}
    \begin{aligned} 
    J_{\pi}(\phi)=\mathbb{E}_{\mathbf{x}_{t} \sim\mathcal{D}}\Bigl[\mathbb{E}_{\mathbf{a}_{t} \sim \pi_{\phi}}\bigl[
    & \alpha \log \left(\pi_{\phi}\left(\mathbf{a}_{t} \mid \mathbf{x}_{t}\right)\right)- \\
    & Q_{\theta}\left(\mathbf{x}_{t}, \mathbf{a}_{t}\right)\bigr]\Bigr]
    \end{aligned}
\end{equation}

A neural network transformation is used to parameterize the policy as $\mathbf{a}_{t}=f_{\phi}\left(\epsilon_{t} ; \mathbf{x}_{t}\right)$, where $\epsilon_{t}$ is an input noise vector sampled from a Gaussian distribution. The objective stated by Eq.~\eqref{eq:J_Pi} can be rewritten as:
\begin{equation}
    \label{eq:J_Pi_v2}
    \begin{aligned} 
    J_{\pi}(\phi)=\mathbb{E}_{\mathbf{x}_{t} \sim \mathcal{D}, \epsilon_{t} \sim \mathcal{N}}\bigl[
    & \alpha \log \pi_{\phi}\left(f_{\phi}\left(\epsilon_{t} ; \mathbf{x}_{t}\right) \mid \mathbf{x}_{t}\right)
    \\
    & -Q_{\theta}\left(\mathbf{x}_{t}, f_{\phi}\left(\epsilon_{t} ; \mathbf{x}_{t}\right)\right)\bigr],
    \end{aligned}
\end{equation}
where $\pi_{\phi}$ is defined implicitly in terms of $f_{\phi}$. The gradient of Eq.~\eqref{eq:J_Pi_v2} is approximated with
\begin{equation}
    \label{eq:J_Pi_grad}
    \begin{aligned} 
    \hat{\nabla}_{\phi} J_{\pi}(\phi)=
    & \nabla_{\phi} \alpha \log \left(\pi_{\phi}\left(\mathbf{a}_{t} \mid \mathbf{x}_{t}\right)\right)+ 
    \bigl(\nabla_{\mathbf{a}_{t}} \alpha \log \left(\pi_{\phi}\left(\mathbf{a}_{t} \mid \mathbf{x}_{t}\right)\right) \\
    & -\nabla_{\mathbf{a}_{t}} Q\left(\mathbf{x}_{t}, \mathbf{a}_{t}\right)\bigr) \nabla_{\phi} f_{\phi}\left(\epsilon_{t} ; \mathbf{x}_{t}\right),
    \end{aligned}
\end{equation}
where $\mathbf{a}_{t}$ is evaluated using $f_{\phi}\left(\epsilon_{t} ; \mathbf{x}_{t}\right)$. 

\noindent \textbf{Remark}: In the maximum entropy framework, the soft policy iteration that alternates between the policy evaluation Eq.~\eqref{eq:J_Q} and the policy improvement Eq.~\eqref{eq:J_Pi} converges to the optimal policy. \textit{Proof in} \cite{haarnoja2018soft}.

\subsection{Action Quantization and Correction}
In the context of SAC learning, the output at any given time $t$ corresponds to the SAC action $\mathbf{a}_t$, which seeks to maximize the policy value $\pi_{\phi}\left(\mathbf{a}_{t} \mid \mathbf{x}_{t}\right)$ with respect to the current SAC state $\mathbf{x}_t$. In order to assess the reward and update the cache, it is necessary to \textbf{\textit{discretize the continuous SAC action}} $\mathbf{a}_t$ and obtain a discrete action $\left(c_{A(t)},\textbf{b},\Delta\textbf{s}\right)$. To achieve this goal, a simple action quantization approach was implemented that relies on thresholding and integer projection.

\textbf{Action quantization}: 
Let us consider an element $\bar{\eta}$ in the SAC action $\mathbf{a}$ and its corresponding quantized version $\eta$ with the selection set $S_{\eta}$. To obtain $\eta$ from $\bar{\eta}$, we adopt a uniform thresholding method for integer projection. Specifically, we use the following equation:
\begin{equation}
\label{eq:action_quan}
\eta = \min{S_{\eta}} + (\bar{\eta}-\min{S_{\eta}}) \, \text{mod} \, \frac{\max{S_{\eta}}-\min{S_{\eta}}}{\max{S_{\eta}}-\min{S_{\eta}} + 1}
\end{equation}

As an example, consider the push action $b_f(t)\in S_{b_f}=\{0,1\}$, we can determine its quantized value $b_f(t) = \bar{b}_f (t) \mod 0.5$ using Eq.~\eqref{eq:action_quan}.

\textbf{Action correction}:
The valid action space of the system is highly constrained due to the limitations imposed by Eq.~\eqref{eq:reacom}, \eqref{eq:reactivebandwidth}, \eqref{cachinp1}, \eqref{cachoutp1}, and \eqref{eq:caupsize}, resulting in a sparsely-spanning space with a cardinality of $(M+1)\times 2^F \times 3^{2F}$. Consequently, even with techniques such as penalty reward, it becomes challenging for the SAC algorithm to identify which actions are valid in this vast space. Therefore, the post-quantization action $\left(c_{A(t)},\textbf{b},\Delta\textbf{s}\right)$ obtained from SAC is often invalid. In order to address this issue, we propose \textit{Rules 1, 5, and 7} to ensure that the output action of SAC is valid, while satisfying the constraints outlined in Section~\ref{sec:sys_action}. Additionally, we introduce \textit{Rules 2, 3, 4, and 6} to improve the training process and enhance the system's overall performance by further compressing the action space, reducing unnecessary costs, and minimizing waste.

\begin{itemize}
\item \textbf{\textit{Rule 1}}: 
When $S^O_{A(t)}$ equals 0, the system checks if the suggested number of computation cores, denoted as $c_{A(t)}$, is less than the minimum workable value given by $\lceil I_{A(t)} w_{A(t)} / (\tau f_D) \rceil$ where $\lceil \cdot \rceil$ represents rounding up to the nearest integer. If this is the case, $c_{A(t)}$ is updated to $\lceil I_{A(t)} w_{A(t)} / (\tau f_D) \rceil$. On the other hand, if $S^O_{A(t)}$ equals 1, $c_{A(t)}$ is set to 0. These rules are designed to fulfill the service latency constraint and reduce unnecessary computation.

\item \textbf{\textit{Rule 2}}: 
When $S^I_f + S^O_f \ge 1$, we set $b_f=0$. This rule indicates that there is no need for proactive pushing if any data of a task is already cached.

\item \textbf{\textit{Rule 3}}: 
To minimize the cost of pushing data to the mobile device, we ensure that at most one task is proactively transmitted, and this task must have the largest $\bar{b}_f$ value among all un-pushed tasks. The selected task will have a $b_f$ value of 1, while all other tasks will have a $b_f$ value of 0. This approach is adopted to avoid unnecessary pushing costs, as the mobile device is only capable of processing one task request per time slot.

\item \textbf{\textit{Rule 4}}: If $b_f=1$, we set $\Delta s^I_f=1$, indicating that the data being proactively pushed needs to be cached.

\item \textbf{\textit{Rule 5}}: 
If the sum of cache sizes given by Eq.~\eqref{eq:caupsize} exceeds the cache capacity, we drop the input or output cache depending on the ascending order of their corresponding $\bar{\textbf{s}}$ values until the cache capacity is satisfied.

\item \textbf{\textit{Rule 6}}: 
If the sum of the caches given by Eq.~\eqref{eq:caupsize} is less than the capacity, we attempt to add reactive input or output cache based on the decreasing order of the continuous variables $\Delta \bar{s}^I_{A(t)}$ and $\Delta \bar{s}^O_{A(t)}$.

\item \textbf{\textit{Rule 7}}: The cache action $\Delta \textbf{s}$ should be clipped according to the minimum and maximum limits specified in Eq.~\eqref{cachinp1} and Eq.~\eqref{cachoutp1}.
\end{itemize}

\subsection{Reward Design}
The reward function $r(\mathbf{x}, \mathbf{a})$ for the SAC state $\mathbf{x}$ and action $\mathbf{a}$ is defined as a function of the resulting bandwidth and computation cost. Specifically, it is given by
\begin{equation}
r(\mathbf{x}, \mathbf{a}) = - \kappa (B(t)+\lambda E(t))
\label{eq:reward}
\end{equation}
where $\kappa$ is a normalization coefficient that is set to $10^{-6}$ in this paper.

The complete SAC learning algorithm is presented in Algorithm~\ref{alg:alg1}. The step sizes for stochastic gradient descent, $\lambda_Q$, and $\lambda_\pi$ are set to $1\times10^{-4}$. The target smoothing coefficient, $\xi$, is chosen to be 0.005.

\begin{algorithm}[h]
\caption{SAC Learning for Our Problem }\label{alg:alg1}
\begin{algorithmic}
\STATE 
\STATE Initialize parameters $\theta, \bar\theta, \phi$ for networks $Q_{\theta}$, $Q_{\bar\theta}$, $\pi_{\phi}$.
\STATE Initialize learning rate $\lambda_Q$, $\lambda_\pi$, and weight $\xi$.
\STATE \textbf{for} each iteration \textbf{do}
\STATE \hspace{0.5cm} \textbf{for} each environment step \textbf{do}
\STATE \hspace{1.0cm} $\mathbf{a}_t \sim \pi_\phi\left(\mathbf{a}_t \mid \mathbf{x}_t\right)$
\STATE \hspace{1.0cm} $\mathbf{x}_{t+1} \sim p\left(\mathbf{x}_{t+1} \mid \mathbf{x}_t, \mathbf{a}_t\right)$
\STATE \hspace{1.0cm} $\mathbf{a}_t$ quantization \& correction, $r\left(\mathbf{x}_t, \mathbf{a}_t\right)$ calculation
\STATE \hspace{1.0cm} $\mathcal{D} \leftarrow \mathcal{D} \cup\left\{\left(\mathbf{x}_t, \mathbf{a}_t, r\left(\mathbf{x}_t, \mathbf{a}_t\right), \mathbf{x}_{t+1}\right)\right\}$
\STATE \hspace{0.5cm} \textbf{end for}

\STATE \hspace{0.5cm} \textbf{for} each gradient step \textbf{do}
\STATE \hspace{1.0cm} $\theta_i \leftarrow \theta_i-\lambda_Q \hat{\nabla}_{\theta_i} J_Q\left(\theta_i\right) \text { for } i \in\{1,2\}$
\STATE \hspace{1.0cm} $\phi \leftarrow \phi-\lambda_\pi \hat{\nabla}_\phi J_\pi(\phi)$
\STATE \hspace{1.0cm} $\bar \theta_i \leftarrow \xi \theta_i + (1-\xi) \bar \theta_i \text { for } i \in\{1,2\}$
\STATE \hspace{0.5cm} \textbf{end for}
\STATE \textbf{end for}

\end{algorithmic}
\end{algorithm}



\section{Implementation and Evaluation}
\subsection{Baselines}
The proposed system is built on the \textbf{\textit{proactive transmission and dynamic-computing-frequency reactive service with cache}}, referred to as \textbf{PTDFC}. For comparison, we have selected the following baselines:
\begin{itemize}
      \item \textbf{\textit{Most-recently-used proactive transmission and least-recently-used cache replacement} (MRU-LRU)}: This is a heuristic algorithm \cite{10.1145/505696.505701, sunpush}, where at each time slot, the requested task is reactively served, and the input data of the most-recently-used task is proactively cached. When the cache is full, the input data cache of the least-recently-used task is replaced. \highlighttext{We choose to cache only the input data, excluding the output data (post-calculation), due to the common scenario where output data tends to be larger in size than input data. This size difference makes caching output data less efficient in the heuristic design for the purpose of reducing overall costs.} The number of computing cores being used is fixed at $0.75M$.
    \item \textbf{\textit{Most-frequently-used proactive transmission and least-frequently-used cache replacement} (MFU-LFU)}: This algorithm is similar to the MRU-LRU algorithm, except that the most/least recently used task is replaced with the most/least frequently used task \cite{10.1145/505696.505701, sunpush}.
    \item \textbf{\textit{Dynamic-computing-frequency reactive service with no cache} (DFNC)}: This algorithm provides reactive service to the requested task, where the mobile device first downloads the input data from the MEC server and then computes it to obtain the output data.
    
    \item \textbf{\textit{Dynamic-computing-frequency reactive service with cache} (DFC)}: This algorithm provides reactive service to the requested task, with the option of caching the input and output data into the limited capacity.
\end{itemize}

It is important to note that the DFC, DFNC, and PTDFC algorithms are all implemented with the SAC algorithm, and as a result, we refer to them as `\textit{SAC-enabled algorithms}' in the following analysis.

\subsection{Data Simulation}
\label{sec:data_sim}
In this study, the training and testing data were generated through a simulation process involving the creation of a Markov chain from a set of tasks $\mathcal{F}$. The transit probability of a task $i$ to another randomly selected task $j \in \mathcal{F}\backslash i$ was established as the maximum transition probability, $p_{i, j} = p_{\max}$. For other tasks $k \in \mathcal{F}\backslash j$, the probability $p_{i, k}$ was calculated as $(1-p_{i, j})\frac{|p^\prime_{i, k}|}{\sum_{f \in \mathcal{F}\backslash j}|p^\prime_{i, f}|}$, where $p^\prime_{i, k}$ or $p^\prime_{i, f}$ were randomly sampled from a uniform distribution. The resulting Markov chain represented the request popularity and transition preferences of $F$ tasks. Subsequently, $10^6$ requested tasks were sampled using a frame-by-frame method. To account for the slow movement of mobile devices, the SNR of the communication channel was dynamically changed every 300 epochs, with four possible values: \SI{0.5}{dB}, \SI{1}{dB}, \SI{2}{dB}, and \SI{3}{dB}. The transition between different SNRs was randomized with equal probabilities. The simulation was conducted using default configurations, which included $M=8$, $F=4$, a maximum transition probability of $0.7$, $\lambda=1$, $I_f$ around $16000$ bits with random offset, $O_f$ around $30000$ bits with random offset, $w=800$ cycles/bit, $\tau=0.02$ seconds, $f_D = 1.7\times 10^8$ cycles/s, $\mu= 10^{-19}$, and $C=40000$ bits.

\subsection{Implementation}
\label{sec:imple}
For the purposes of training and stabilization, the SAC action $\mathbf{a}_t$ and system state $\mathbf{x}_t$ are normalized to fall within the range of $[-1, 1]$. Implementation of the system is accomplished through the use of Python and PyTorch. Training and testing processes are executed on a computer with a TITAN RTX GPU, utilizing a batch size of 256, a discount factor of $\gamma=0.99$, automatic entropy temperature $\alpha$ tuning \cite{haarnoja2018soft}, a hidden-layer size of 256, one model update per step, one target update per 1000 steps, and a replay buffer size of $1\times10^7$. The testing process is executed 10 epochs after every 10 training epochs, and the training and testing processes are halted when the reward and loss have converged.

\begin{figure}[!h]
\centering
\includegraphics[width=0.48\textwidth]{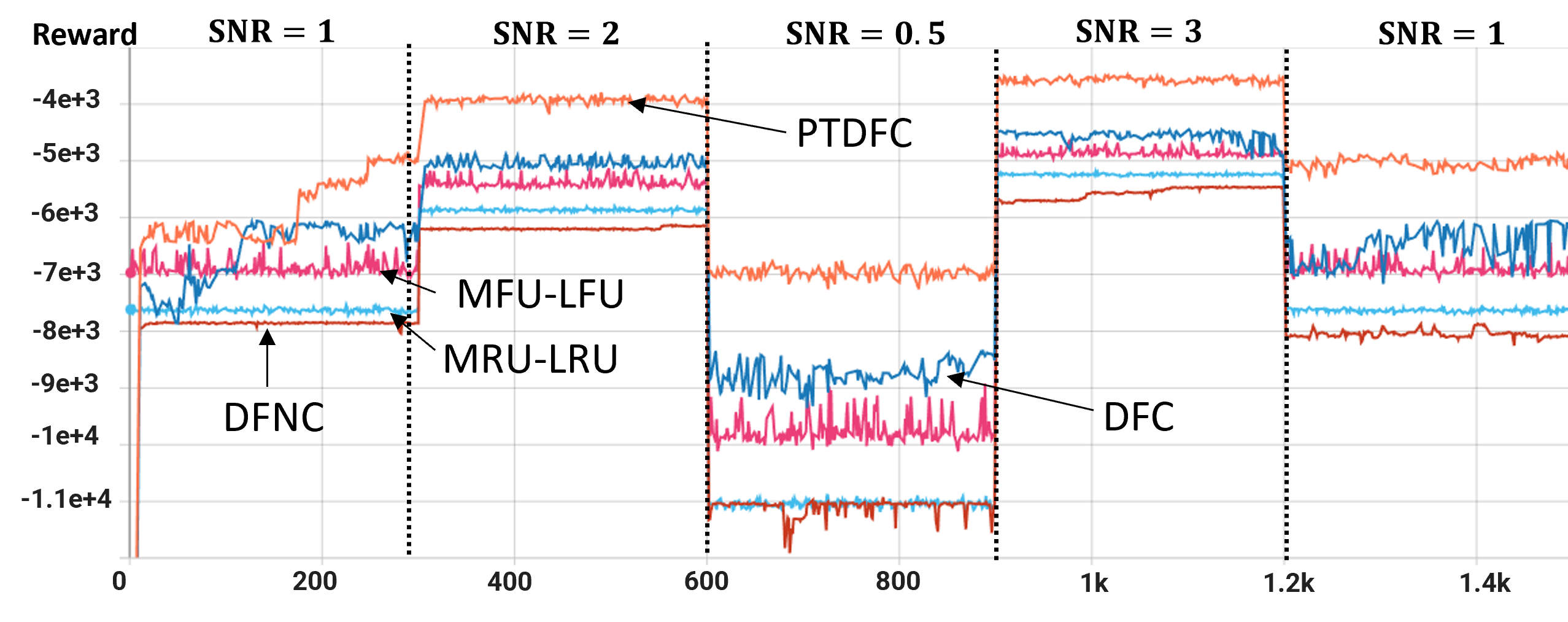}
  \caption{Training reward of PTDFC, DFC, DFNC, MFU-LFU, MRU-LRU algorithms when the SNR values are dynamically changed every 300 epochs.}
  \label{fig:converge}
\end{figure}

\subsection{Convergence Analysis}
We present the training convergence results of three SAC-based algorithms, PTDFC, DFC, DFNC, and two heuristic algorithms, MFU-LFU and MRU-LRU in Fig.~\ref{fig:converge}. The curves plot the reward versus epochs under different SNR conditions for these algorithms. \highlighttext{It is important to note that the MFU-LFU and MRU-LRU algorithms are heuristic in nature, lacking parameters for training. Despite this, we have included their reward outcomes in Fig.~\ref{fig:converge}, aiming to provide a more comprehensive perspective on their relative performance compared to others.} \highlighttext{During the first 300 epochs, with $\text{SNR}=\SI{1}{dB}$, the PTDFC, DFC, and DFNC algorithms commence with neural network parameters initialized randomly and achieve convergence in a substantial number of epochs (250, 135, and 20 epochs, respectively).} PTDFC requires more training epochs to converge than DFC and DFNC simply because of its larger action space. Starting from epoch 300, the SNR value is increased to \SI{2}{dB}, and the three SAC-based algorithms converge again in less than 13 epochs using the pre-trained model from the previous epochs. This finding demonstrates the remarkable generalization ability of SAC-based algorithms to handle SNR change cases. These SAC-based algorithms can get fine-tuned and converged again within a few epochs (around 10). The quick convergence ability is also validated at epochs 600 and 900 when the SNR changes to \SI{0.5}{dB} and \SI{3}{dB}, respectively. \highlighttext{We also noticed that there are significant discrepancies in the convergence time between the two $\text{SNR}=\SI{1}{dB}$ stages. In the later $\text{SNR}=\SI{1}{dB}$ stage (epoch 1200-1500), the PTDFC achieves convergence in approximately 10 epochs after the SNR transition. This can be attributed to the solid foundation of well-trained parameters established during the preceding $\text{SNR}=\SI{3}{dB}$ stage. This trend reaffirms the system's adeptness in rapidly adapting to environmental SNR changes.}


The optimization problem at hand involves both linear and nonlinear objective functions, constraints, and involves binary variables. Consequently, it is classified as an Integer Nonlinear Programming (INLP) problem and is notoriously difficult to solve. Traditional optimization algorithms for INLP (such as Branch and Bound) are not suitable for this problem due to their exponential convergence time and the assumption of global knowledge of the environment and its dynamics. Moreover, in the event of a change in the environment, such as a variation in channel SNR, it takes a considerable amount of time to solve the problem and achieve convergence again. Classical machine learning algorithms, exceptionally standard reinforcement learning, also face challenges in scaling with the large dimension of the variable set, requiring an excessively large network size and convergence time to solve the problem.

\begin{figure}
    \centering  \includegraphics[width=0.5\textwidth, trim=1 3 1 1,clip]{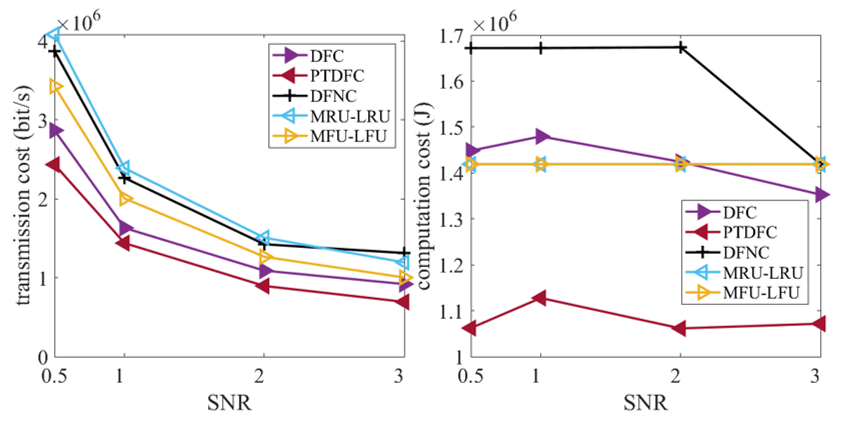}
    \caption{The system performance for the proposed PTDFC algorithm and the baselines with the configuration stated in Section.~\ref{sec:data_sim} and \ref{sec:imple}. (left) Transmission cost vs. SNR. (right) Computation cost vs. SNR.}
  \label{fig:res_dynamic_snr}
\end{figure}

\begin{figure*}[!t]
    \centering
    \includegraphics[width=7in, trim=1 3 1 1,clip]{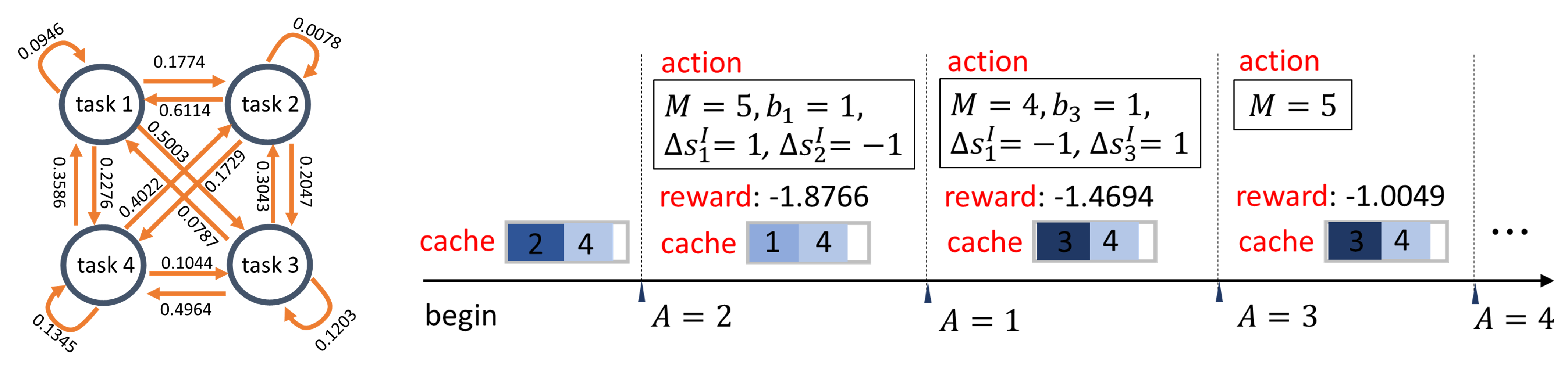}
    \caption{A qualitative example for the joint optimization of 4 tasks using SAC-based PTDFC algorithm. (left) Visualization of the Markov transition probability among 4 tasks. (right) The requested task, cache state, action, and reward for the first four time slots of the proposed SAC system. If no action is mentioned, it defaults to no change with a value of 0.}
  \label{fig:qua}
\end{figure*}

\subsection{Numerical Results}
The system performance of the proposed PTDFC algorithm and the baselines (DFC, DFNC, MRU-LRU, MFU-LFU) in terms of transmission bandwidth cost and computation cost for different channel SNRs is presented in Fig.~\ref{fig:res_dynamic_snr}. The results show that the PTDFC algorithm achieves the lowest cost for transmission bandwidth for various channel SNRs, followed by the DFC, MFU-LFU, DFNC, and MRU-LRU algorithms. In terms of computation cost, the top-performing algorithms are PTDFC, MFU-LFU/MRU-LRU, DFC, and DFNC, respectively. Overall, the PTDFC algorithm achieves a reduction of around $5\times 10^5$ bits/s in transmission cost and $3\times 10^5$ J in computation cost for every SNR condition, compared to the second-best algorithm. It is also observed that all algorithms take less transmission bandwidth for the requested task as the SNR value increases, indicating that a higher SNR results in better channel quality.

\subsection{Qualitative Results Analysis}
Fig.~\ref{fig:qua} provides an example of the status and action of four requests when deploying the PTDFC algorithm. The figure visualizes the requested task, cache state, action, and reward of each time slot to show the joint computing, pushing, and caching optimization of the four tasks. In this example, at $t_0$, the mobile device requests task 1 from the MEC server, which has empty cache content. The system then makes reactive transmission and computing for task 1 with five cores and pushes the input data of unrequested task 3, followed by caching the input data of tasks 1 and 3. At $t_1$, the requested task is task 3, and the system makes the reactive computing of the cached task 3 with four cores and pushes the input data of task 4. Then, the system replaces the cache of task 1 with the input data for task 4. Similarly, at $t_2$, the requested task is task 4, and the system makes the reactive computing of the cached task 4 with five cores and pushes the input data of task 2. Finally, the system removes the cache for task 3 and caches the input data for task 2. The example illustrates that the SAC-based PTDFC system is capable of predicting the user's future requests using deep networks and pushing or caching the appropriate content to enhance system performance.

\subsection{Tuning Analysis}
In this section, we investigate the impact of several crucial parameters on the performance of the proposed PTDFC algorithm. These parameters include the cache size ($C$), number of computing cores ($M$), number of tasks ($F$), maximum transition probabilities, base computing frequency ($f_D$), task input size ($I_f$), task output size ($O_f$), tolerable service delays ($\tau$), and cost weights ($\lambda$). To analyze the effects of each parameter, we hold the other parameters constant and observe the resulting changes in performance. The default values for these parameters are specified in Section~\ref{sec:data_sim}, and we maintain a fixed channel SNR value of 1 to isolate the effects of parameter tuning.

\subsubsection{Different Cache Size $C$}
\begin{figure}[!t]
\centering
\includegraphics[width=0.5\textwidth]{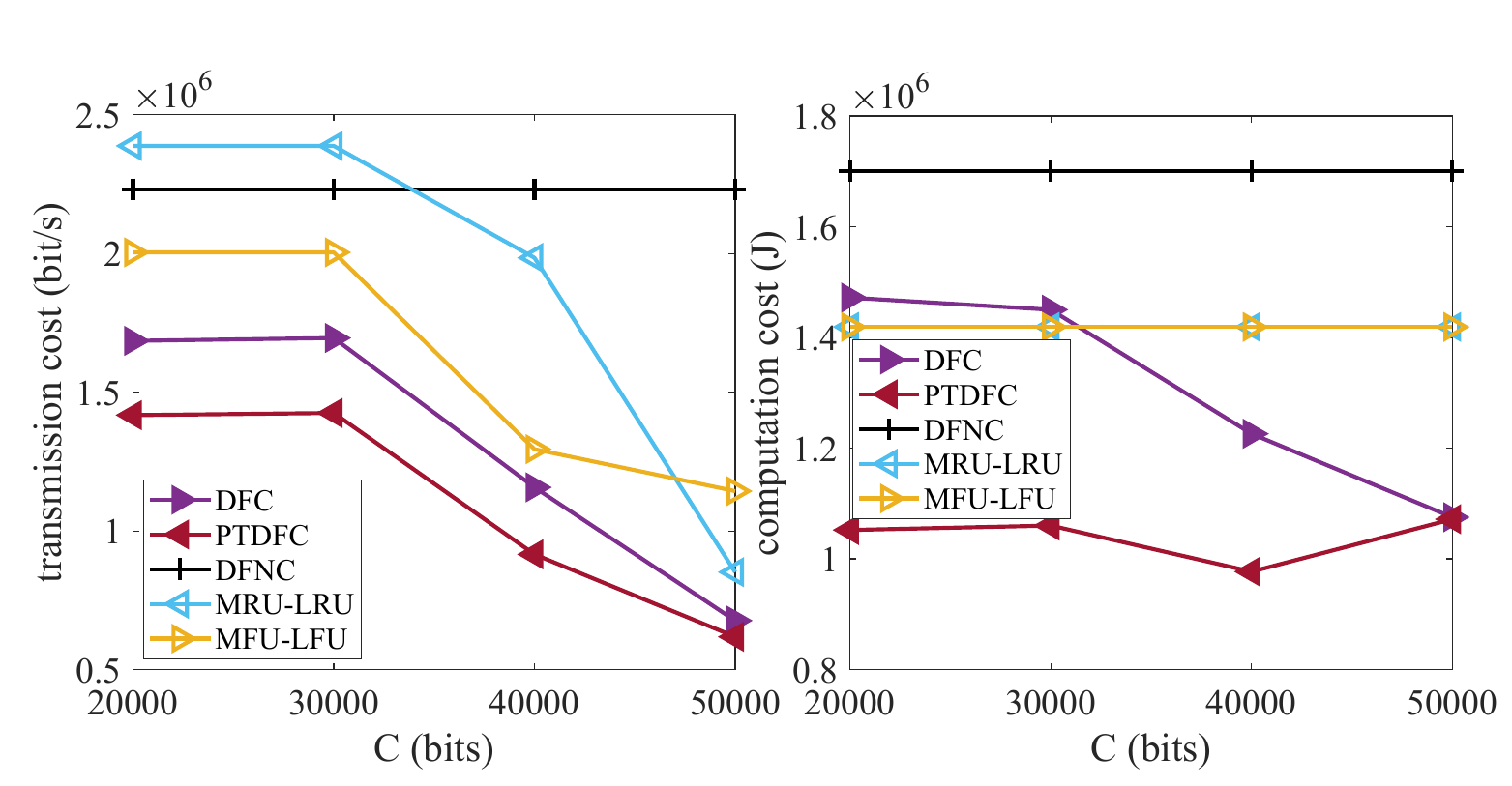}
  \caption{Impact of varying the cache size $C$ when using the default configuration and fixed SNR. (left) Transmission cost vs. $C$. (right) Computation cost vs. $C$.}
  \label{fig:varyC}
\end{figure}

In Fig.~\ref{fig:varyC}, we have presented the averaged transmission and computation costs of three SAC-enabled algorithms, DFNC, DFC, and PTDFC, as well as two heuristic algorithms, MRU-LRU and MFU-LFU, under different cache sizes $C$. It is worth noting that the DFNC algorithm is not affected by changes in cache size, as it only provides reactive service without caching. The other algorithms show a decrease in transmission costs as the cache size is increased, due to the availability of more locally cached input data. Moreover, our proposed PTDFC algorithm consistently achieves lower transmission and computation costs than the other algorithms, thanks to its ability to dynamically adjust the cache via proactive transmission. With a very large cache size, (e.g., $C=50000$ bits), the performance of PTDFC and DFC is similar because the cache is large enough to store all input data and there is no need for a proactive transmission. \highlighttext{Furthermore, we have observed a consistent overlapping trend in the computation costs of MRU-LRU and MFU-LFU across various configurations. This overlapping behavior can be attributed to our design choice in both algorithms, wherein solely the input data is cached. Consequently, the performance of MRU-LRU and MFU-LFU in terms of computation cost tends to align, as governed by Eq.~\eqref{eq:reactiveenergy}, when $S_{A(t)}^O(t)=0$.}

\begin{figure}[!t]
\centering
\includegraphics[width=0.5\textwidth]{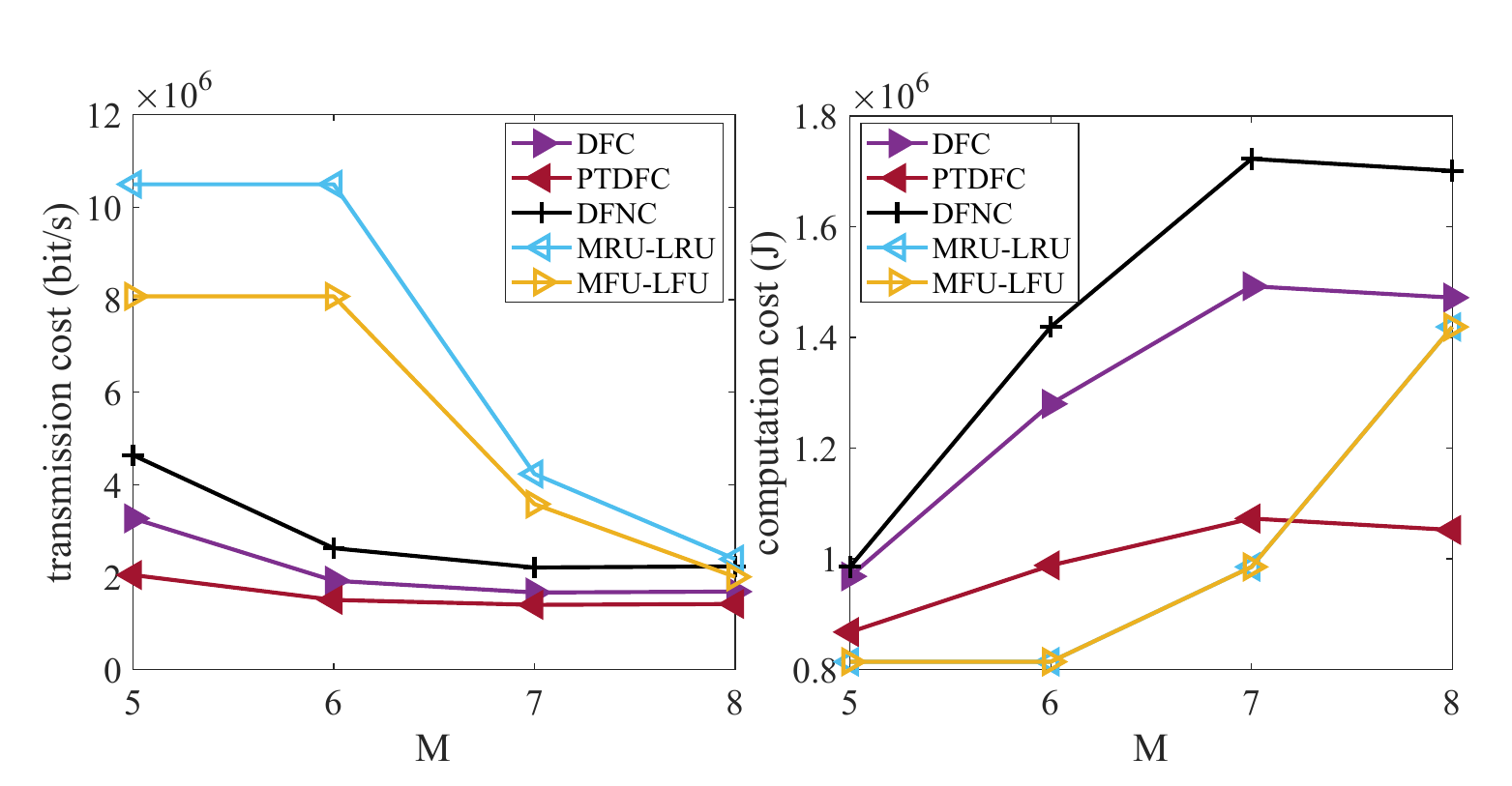}
\caption{Impact of varying the number of computing cores $M$ when using the default configuration and fixed SNR. (left) Transmission cost vs. $M$. (right) Computation cost vs. $M$.}
  \label{fig:varyM}
\end{figure}

\subsubsection{Different Number of Computation Cores $M$}
Fig.~\ref{fig:varyM} illustrates the performance of five algorithms, namely DFNC, DFC, PTDFC, MRU-LRU, and MFU-LFU, under different numbers of computation cores $M$. As the number of computing cores increases, the transmission cost of all five algorithms decreases, while their computation cost increases correspondingly. This is because all the reactive tasks are time-sensitive, and the system tends to utilize more computing cores for computing to reduce the computing time and leave more time for reactive transmission, which would effectively reduce the transmission cost and total cost. The proposed PTDFC algorithm consistently achieves a low transmission cost and computation cost by selecting an appropriate computing core number for the required task to achieve a better reward or a smaller cost. In contrast, the heuristic algorithms MRU-LRU and MFU-LFU have high transmission costs for small $M$ and significant computing costs for large $M$, as their computing frequency is linearly adjusted with the increase of computing cores.

\subsubsection{Different Number of Tasks $F$}
\begin{figure}[!t]
\centering
\includegraphics[width=0.5\textwidth]{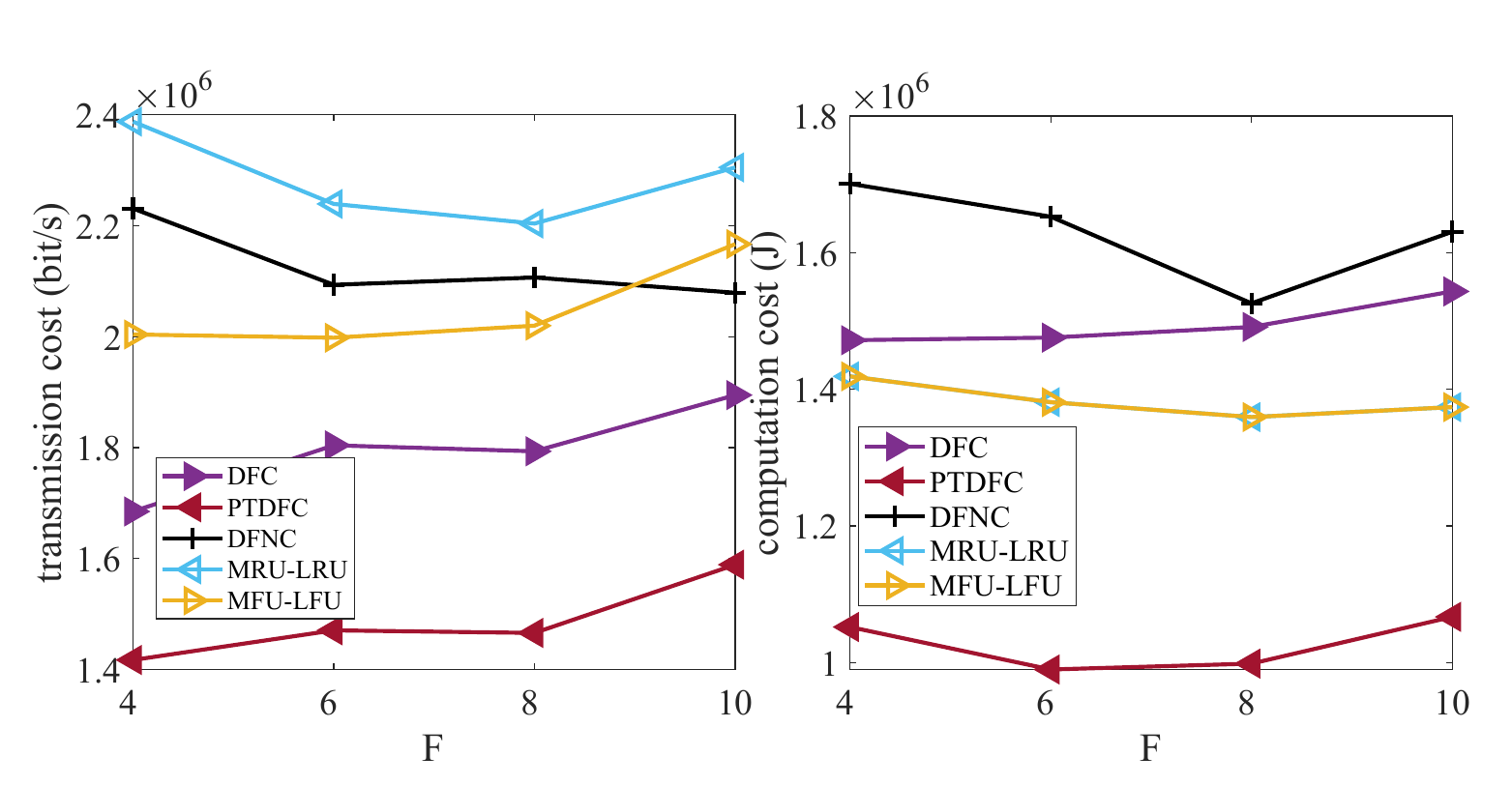}
  \caption{Impact of varying the task number $F$ when using the default configuration and fixed SNR. (left) Transmission cost vs. $F$. (right) Computation cost vs. $F$.}
  \label{fig:varyF}
\end{figure}

In Fig.~\ref{fig:varyF}, we present the performance of five algorithms under various numbers of tasks in the request set. As the number of tasks increases, we observe a slight increase in the transmission and computation costs for the DFC and PTDFC algorithms, possibly because only a small portion of all tasks can be cached or proactively transmitted, while the rest has to be reactively served, leading to higher costs. However, it is worth noting that the proposed PTDFC algorithm consistently outperforms all other algorithms across different numbers of tasks. This is due to its ability to dynamically select the best task in the task set for caching and proactive transmission, thereby minimizing costs associated with reactive service.

\begin{figure}[!t]
\centering
\includegraphics[width=0.5\textwidth]{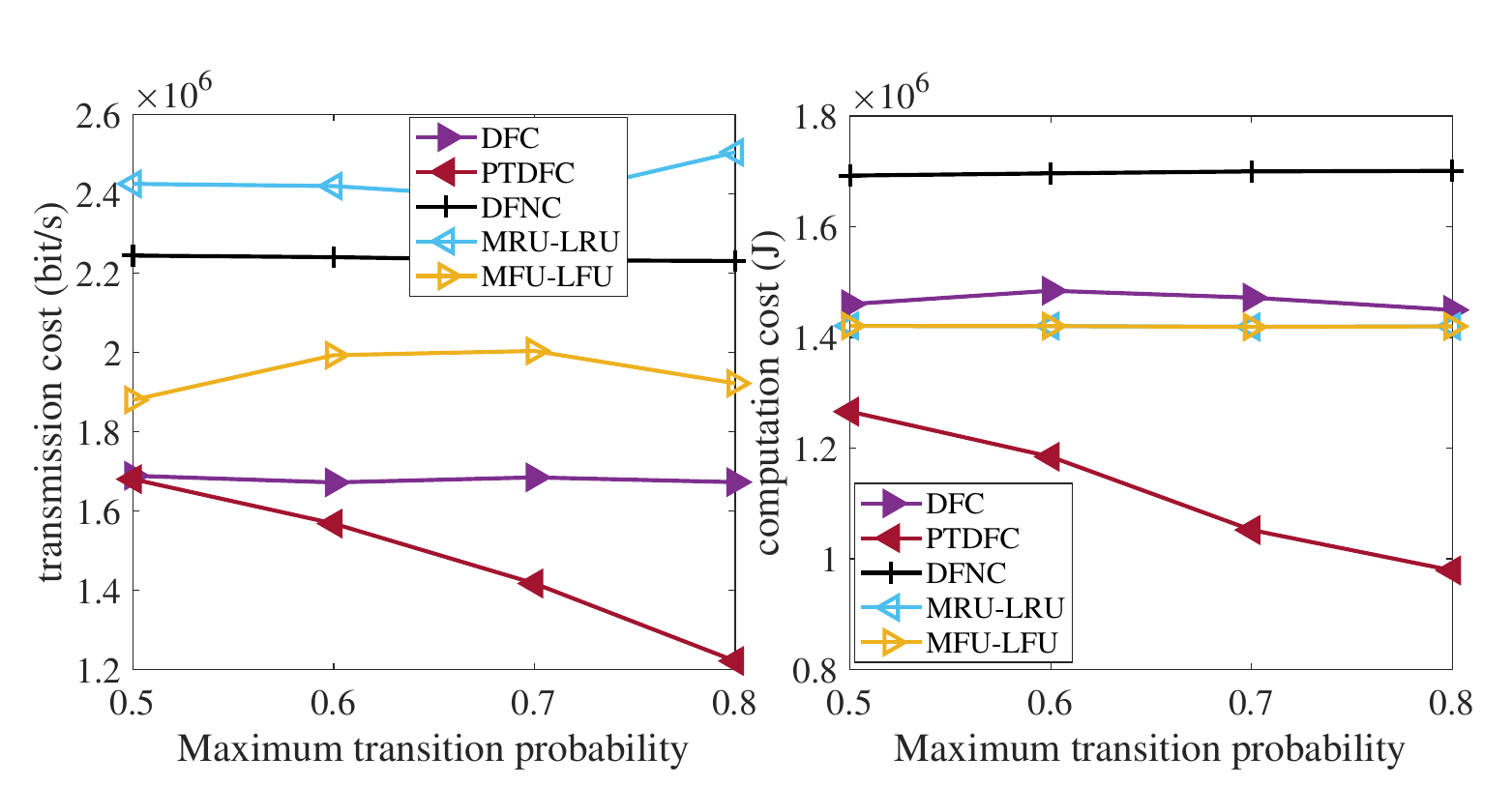}
  \caption{Impact of varying the maximum transition probability in Markov chain when using the default configuration and fixed SNR. (left) Transmission cost vs. maximum transition probability. (right) Computation cost vs. maximum transition probability.}
  \label{fig:varyTP}
\end{figure}

\subsubsection{Different Maximum Transition Probabilities} 
We investigated the impact of the maximum transition probabilities $p_{\max}$ on the Markov chain simulation, which is a crucial parameter. Accordingly, we conducted experiments by varying the $p_{\max}$ value and evaluated the performance of five algorithms under different $p_{\max}$ values. The results, shown in Fig.~\ref{fig:varyTP}, indicate that a higher $p_{\max}$ value leads to an easier prediction of future service requests based on the current request, resulting in the corresponding proactive push operation. Consequently, the PTDFC algorithm, equipped with proactive transmission, demonstrated significant reductions in transmission and computation costs compared to the other algorithms, which do not consider proactive transmission. It is worth noting that the other algorithms exhibited little to no variation in their performance across different $p_{\max}$ values, as they do not rely on the push operation.

\subsubsection{Different Base Computing Frequency $f_D$}
\begin{figure}[!t]
\centering
\includegraphics[width=0.5\textwidth]{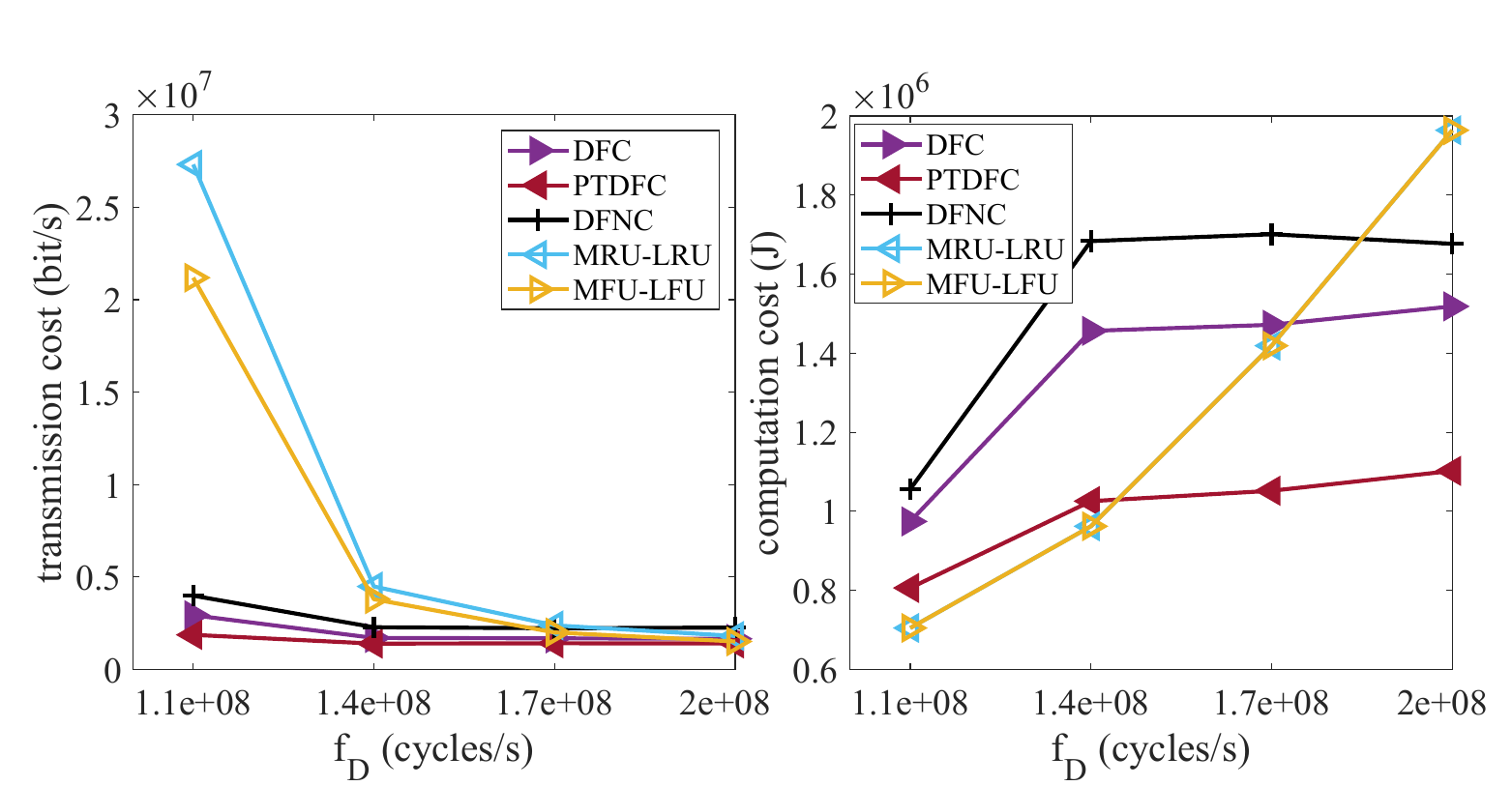}
  \caption{Impact of varying the base computing frequency $f_D$ when using the default configuration and fixed SNR. (left) Transmission cost vs. $f_D$. (right) Computation cost vs. $f_D$.}
  \label{fig:varyfD}
\end{figure}

To study the impact of computing frequency on system performance, we conducted an evaluation of five algorithms with varying $f_D$ and presented our results in Fig.~\ref{fig:varyfD}. As a general rule, increasing the computing frequency is equivalent to having more computing cores with a fixed value of $f_D$. Therefore, the trends observed in the results of Fig.~\ref{fig:varyfD} are similar to those previously reported in Fig.~\ref{fig:varyM}. Notably, the PTDFC algorithm consistently exhibits the lowest overall cost (i.e., the sum of transmission and computation costs) across all $f_D$ configurations.

\subsubsection{Different Task Input Size $I_f$}
\begin{figure}[!t]
\centering
\includegraphics[width=0.5\textwidth]{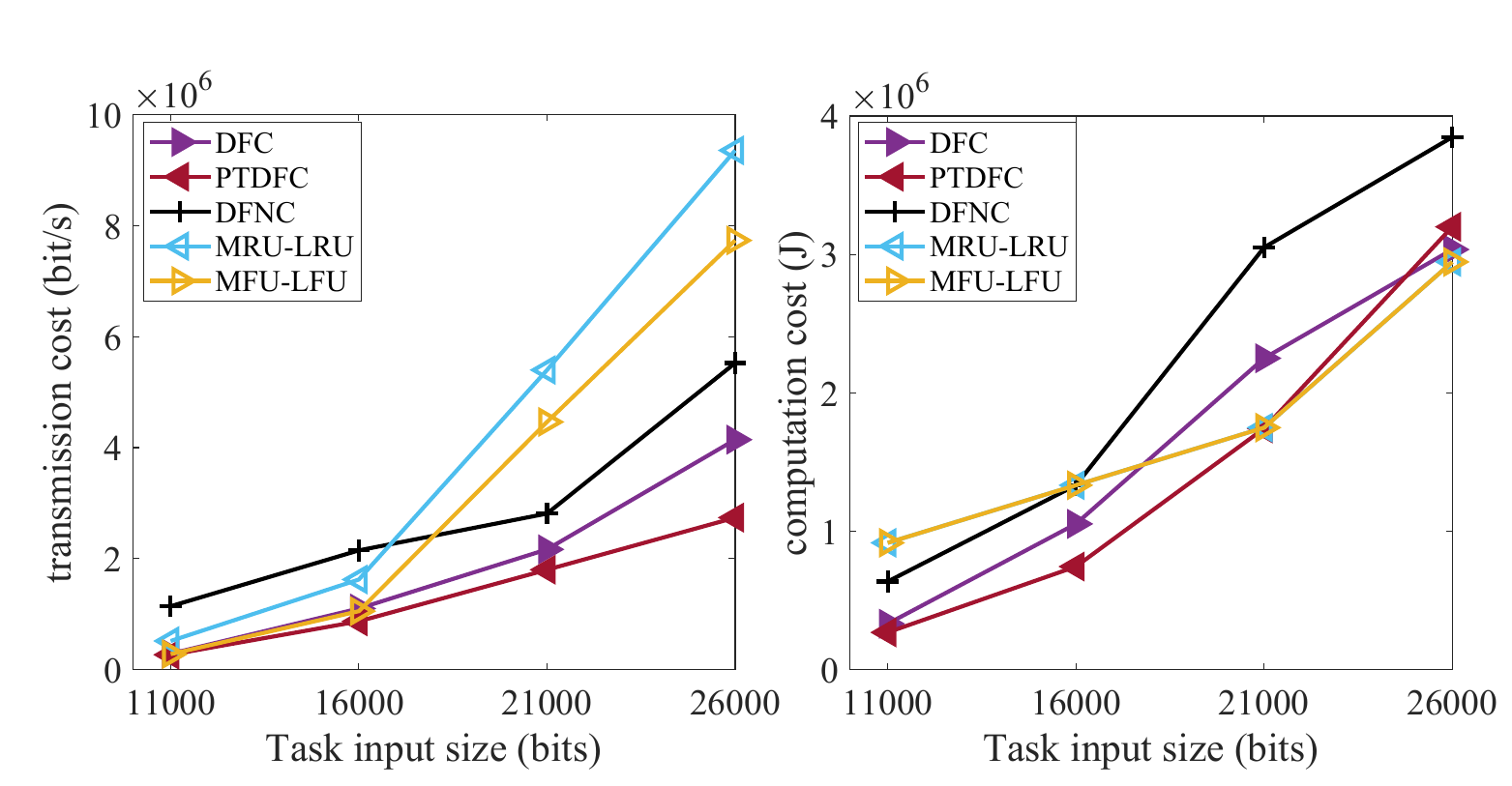}
  \caption{Impact of varying the input data size $I_f$ for 4 tasks when using the default configuration and fixed SNR. (left) Transmission cost vs. $I_f$. (right) Computation cost vs. $I_f$.}
  \label{fig:varyIS}
\end{figure}

The default input data size for all tasks is $16000$ bits. To investigate the impact of input data size on the system performance, we conducted experiments by varying the input data size for four tasks and evaluated the performance of five algorithms under different $I_f$.  The results are presented in Fig.~\ref{fig:varyIS}, where it can be observed that an increase in the input data size leads to an increase in both transmission and computation costs for all algorithms. This can be attributed to the fact that larger input data requires more bandwidth and computation, whether in reactive or proactive cases, as indicated by Eq.~\eqref{eq:reactivebandwidth}, \eqref{eq:reactiveenergy}, \eqref{eq:probandwidth}. On the other hand, when the input data size is relatively small (e.g., 11000 bits), the PTDFC and DFC algorithms exhibit similar performance, as the optimal policy for both algorithms is to cache as much input data as possible and provide full reactive service only for non-cached tasks. However, in general, the PTDFC algorithm consistently outperforms the other algorithms, achieving the smallest total cost across all input data size configurations.

\subsubsection{Different Task Output Size $O_f$} 
\begin{figure}[!t]
\centering
\includegraphics[width=0.5\textwidth]{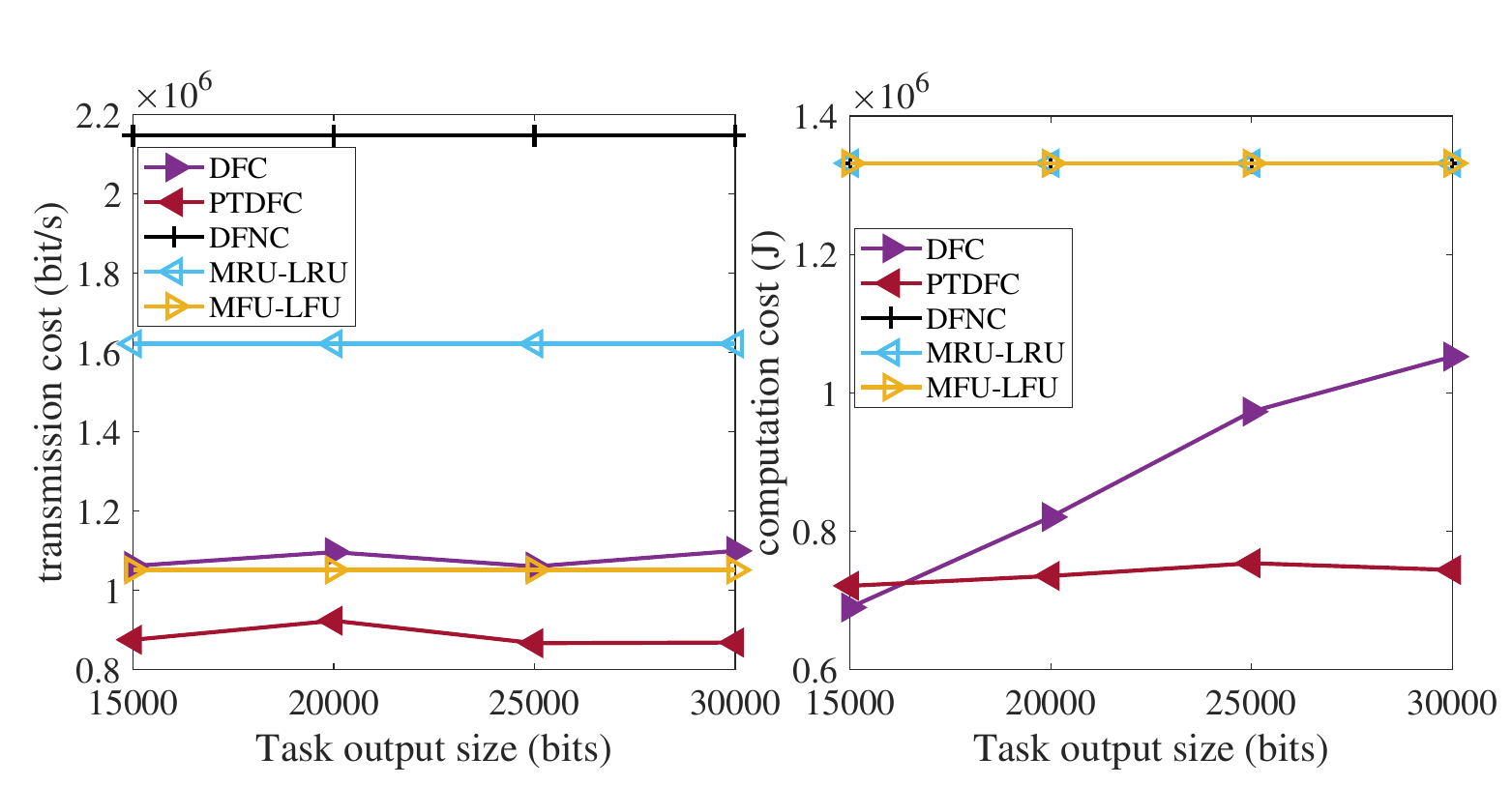}
  \caption{Impact of varying the output data size $O_f$ for 4 tasks when using the default configuration and fixed SNR. (left) Transmission cost vs. $O_f$. (right) Computation cost vs. $O_f$.}
  \label{fig:varyOS}
\end{figure}

After analyzing the impact of input data size, we further examined the influence of output data size on the system performance. For this purpose, we altered the output data size of four tasks and evaluated the performance of five algorithms under different $O_f$. The default output data size for all tasks was set to $30000$ bits. As depicted in Fig.~\ref{fig:varyOS}, the transmission cost and computation cost for the DFNC, MRU-LRU, and MFU-LFU algorithms remained unchanged, as the output data size did not affect the calculation of the two costs and the cache update mechanism. However, for the DFC algorithm, the transmission cost fluctuated around a constant level, and the computation cost increased as the output data size increased from $15000$ bits to $30000$ bits. This behavior can be attributed to the algorithm's prioritization of caching input data, which ensures a low transmission cost. On the other hand, due to limited cache size, lower priority, and increased output data size, only a small fraction of tasks had the opportunity to cache their output data, resulting in additional computation for reactive service. In contrast, the PTDFC algorithm's joint computing, pushing, and caching design demonstrated robustness to variations in $O_f$, as evidenced by the flat transmission and computation cost curves in Fig.~\ref{fig:varyOS}.

\subsubsection{Different Tolerable Service Delays $\tau$} 
\begin{figure}[!t]
\centering
\includegraphics[width=0.5\textwidth]{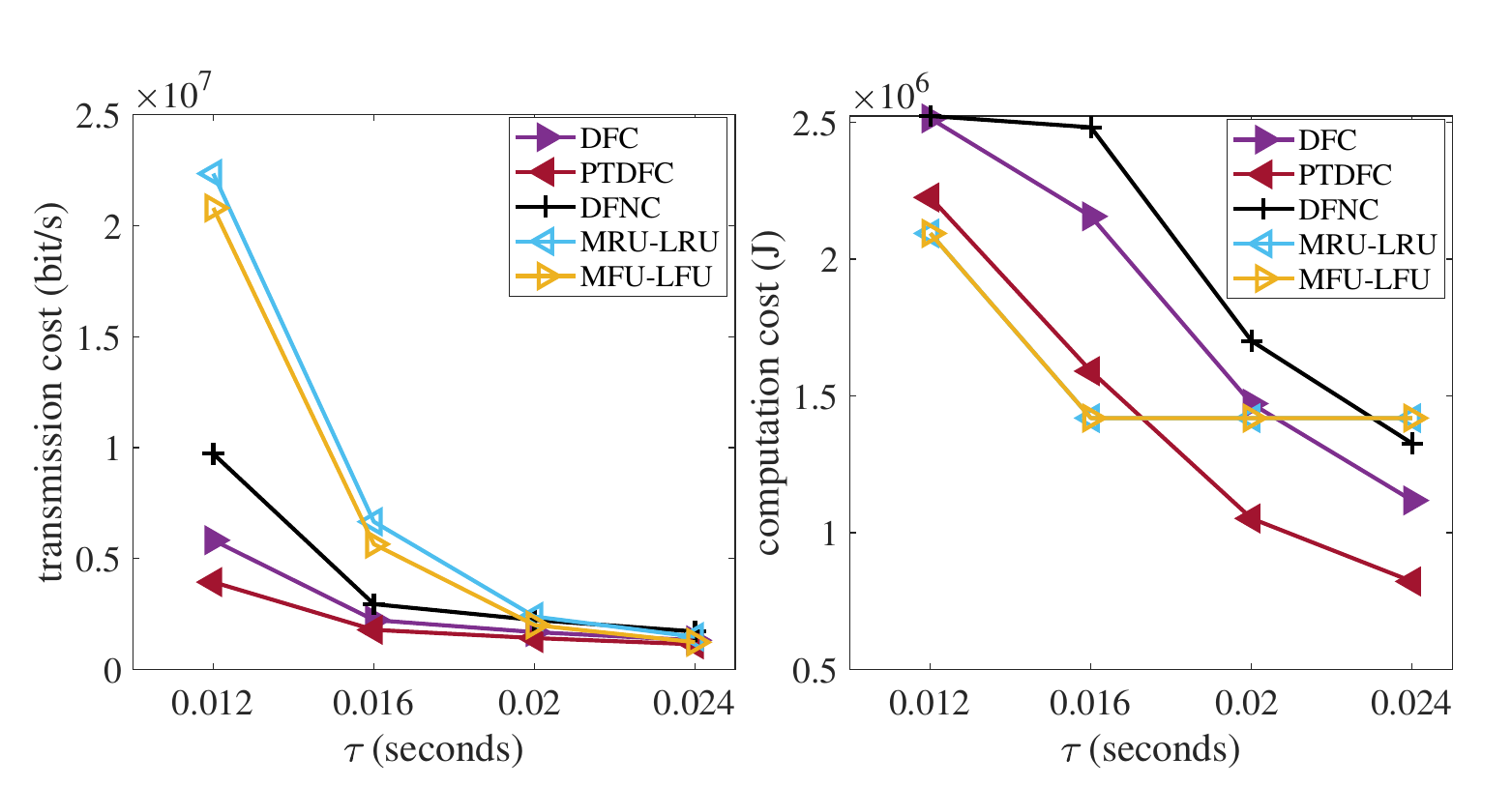}
  \caption{Impact of varying the maximum tolerable service latency $\tau$ when using the default configuration and fixed SNR. (left) Transmission cost vs. $\tau$. (right) Computation cost vs. $\tau$.}
  \label{fig:varyTau}
\end{figure}

The tolerable service delay is a critical parameter that significantly influences the transmission and computation costs of the system. To evaluate their impact, we tested the performance of five algorithms under varying values of $\tau$ and present our findings in Fig.~\ref{fig:varyTau}. As we increase $\tau$ from \SI{0.012}{s} to \SI{0.024}{s}, we observe a corresponding decline in the transmission and computation costs for most algorithms. This is because the larger $\tau$ provides more time for transmitting the input data of the requested task, reducing the bandwidth cost as per Eq.~\eqref{eq:reactivebandwidth}. Similarly, additional processing time is given to the computation step for acquiring the output data, relaxing the requirement of the computing frequency, and leading to a lower computation cost as per Eq.~\eqref{eq:reactiveenergy}. Notably, the PTDFC algorithm outperforms the other algorithms by achieving the lowest transmission and computation cost under all $\tau$ values. This is attributed to its ability to design an optimal policy for joint computing, pushing, and caching through deep reinforcement learning. However, the transmission cost of all five algorithms converges at larger $\tau$, and the benefits of the PTDFC algorithm are mitigated as a lower computing frequency (one core) is employed for all algorithms, resulting in comparable transmission costs.

\begin{figure}
\centering
\includegraphics[width=0.5\textwidth]{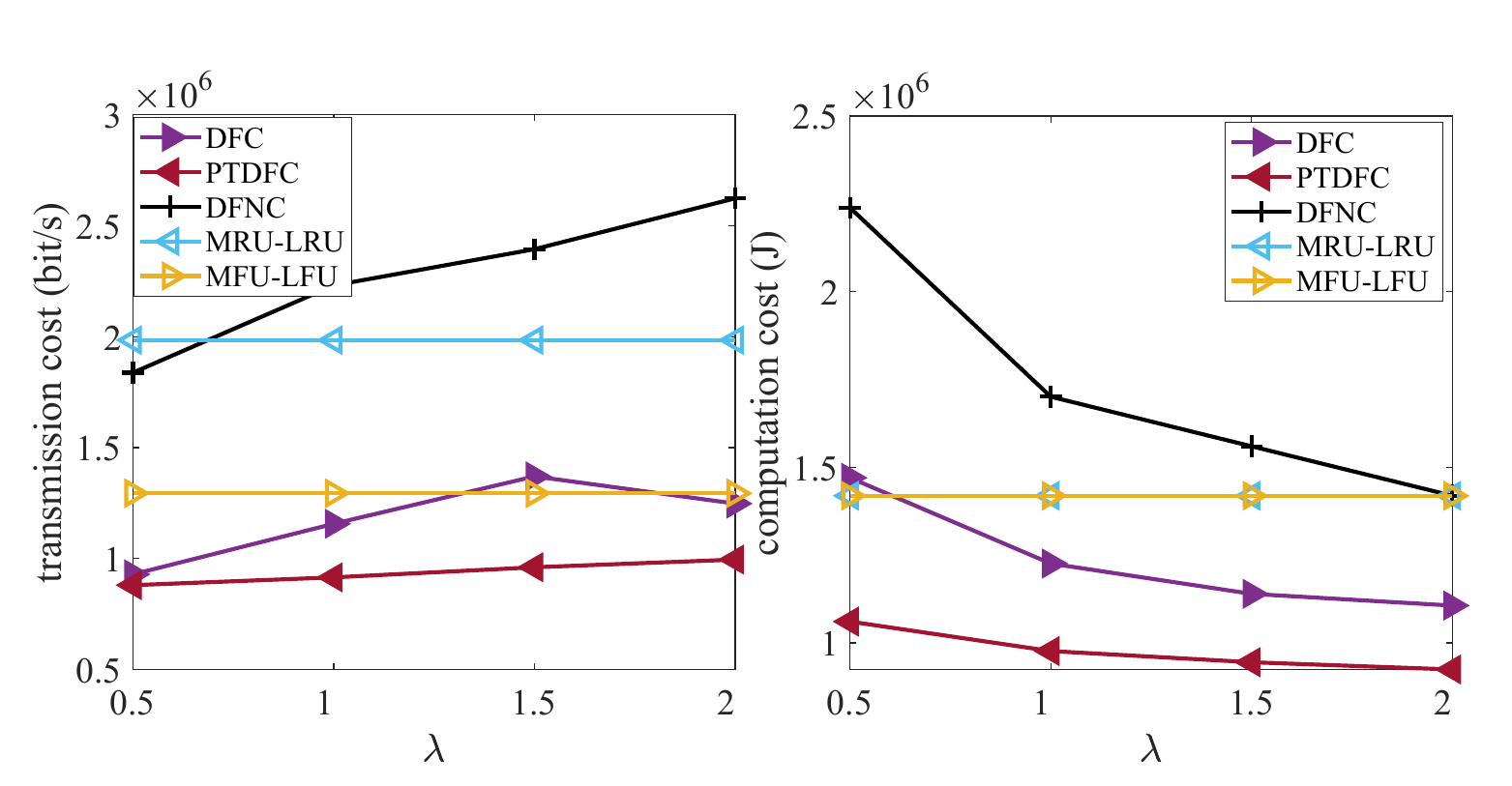}
  \caption{Impact of varying cost weight $\lambda$ when using the default configuration and fixed SNR. (left) Transmission cost vs. $\lambda$. (right) Computation cost vs. $\lambda$.}
  \label{fig:varyLambda}
\end{figure}

\subsubsection{Different Cost Weights $\lambda$}
To guide policy learning for the trade-off between transmission cost and computation cost, the default cost weight of 1 is used for designing the reward function in Eq.~\eqref{eq:reward}. To investigate the impact of this parameter on the performance of the algorithms, we evaluated the performance of five algorithms under different $\lambda$ values and present the results in Fig.~\ref{fig:varyLambda}. Notably, the SAC-enabled algorithms show a clear trade-off between the two costs, as increasing $\lambda$ leads to a decrease in computation cost and a corresponding increase in transmission cost, while the heuristic algorithms MRU-LRU and MFU-LFU have completely flat cost curves. The PTDFC algorithm consistently achieves the best performance under different $\lambda$ values, which can be attributed to its optimal computing, pushing, and caching policy design through deep reinforcement learning.

\subsection{Complexity Analysis}
\highlighttext{The computational complexity of the proposed PTDFC algorithm largely relies on the number and structure of neural networks in SAC system \cite{9760005}. At the training stage, Algorithm~\ref{alg:alg1} incorporates the parameter updating of three neural networks: $Q_{\theta}$, $Q_{\bar\theta}$ (the actors), and $\pi_{\phi}$ (the critic). Therefore, the computation of the complexity of Algorithm~\ref{alg:alg1} is:}
\begin{equation}
\begin{aligned}
& 2 \times \sum_{j=0}^{J-1} n_j^{Q} n_{j+1}^{Q}+ \times \sum_{k=0}^{K-1} n_k^{\pi} n_{k+1}^{\pi} \\
&=O\left(\sum_{j=0}^{J-1} n_j^{Q} n_{j+1}^{Q}+\sum_{k=0}^{K-1} n_k^{\pi} n_{k+1}^{\pi}\right)
\end{aligned}
\end{equation}
\highlighttext{where $J$ denotes the number of fully connected layers for the $Q_{\theta}$ and $Q_{\bar\theta}$ networks (having identical structure), and $K$ denotes that for $\pi_{\phi}$ network. $n_j^{Q}$ and $n_k^{\pi}$ represent the number of neurons at the $j$-th layer of $Q_{\theta}$ or $Q_{\bar\theta}$ networks and the $k$-th layer of $\pi_{\phi}$ network. $j = 0$ and $k = 0$ represent the input layers.} 

\highlighttext{At the testing stage, Algorithm~\ref{alg:alg1} only needs to execute the trained $Q_{\theta}$, $Q_{\bar\theta}$ networks, so the computation complexity is reduced to $O\left(\sum_{j=0}^{J-1} n_j^{Q} n_{j+1}^{Q}\right)$. In our system, $J=3$, $K=4$, $N_j^Q= 22, 256, 256, 1$ for $j=0,1,2,3$, given the number of tasks $F=4$.}

\balance
\section{Conclusion}
In this paper, we explore joint optimization of computing, pushing, and caching in MEC networks to further improve user-perceived quality of experience. We formulate the joint-design problem as an infinite-horizon discounted-cost Markov decision process, which allows us to optimize the total quantity of transmitted data and the total computation cost for the mobile user. To solve this problem, we propose a framework based on SAC learning that dynamically orchestrates the three functions. The framework is featured with embedded deep networks that implicitly predict user future requests and a design for action quantization and correction that enables SAC to work for this problem. In simulations using a single-user single-server MEC network, our proposed framework effectively reduces both transmission load and computing cost and outperforms baseline algorithms across various parameters.

\bibliographystyle{IEEEtran}
\bibliography{bibtex}

\end{document}